# Pure Component Property Estimation Framework Using Explainable Machine Learning Methods

Jianfeng Jiao[1], Xi Gao[2,3,§] and Jie Li[1,*]

[1]Centre for Process Integration, Department of Chemical Engineering, School of Engineering, The University of Manchester, Manchester M13 9PL, UK
[2]School of Electronic and Information Engineering, Tongji University, Shanghai, China 201804
[3]School of Mechanical and Electrical Engineering, Jinggangshan University, Ji'an, Jiangxi, China 343009

**Abstract**

Accurate prediction of pure component physiochemical properties is crucial for process integration, multiscale modeling, and optimization. In this work, an enhanced framework for pure component property prediction by using explainable machine learning methods is proposed. In this framework, the molecular representation method based on the connectivity matrix effectively considers atomic bonding relationships to automatically generate features. The supervised machine learning model random forest is applied for feature ranking and pooling. The adjusted $R^2$ is introduced to penalize the inclusion of additional features, providing an assessment of the true contribution of features. The prediction results for normal boiling point ($T_b$), liquid molar volume ($L_{mv}$), critical temperature ($T_c$) and critical pressure ($P_c$) obtained using Artificial Neural Network and Gaussian Process Regression models confirm the accuracy of the molecular representation method. Comparison with GC based models shows that the root-mean-square error on the test set can be reduced by up to 83.8%. To enhance the interpretability of the model, a feature analysis method based on Shapley values is employed to determine the contribution of each feature to the property predictions. The results indicate that using the feature pooling method reduces the number of features from 13316 to 100 without compromising model accuracy. The feature analysis results for $T_b$, $L_{mv}$, $T_c$, and $P_c$ confirms that different molecular properties are influenced by different structural features, aligning with mechanistic interpretations. In conclusion, the proposed framework is demonstrated to be feasible and provides a solid foundation for mixture component reconstruction and process integration modelling.

---

[*] The corresponding author: Jie Li (jie.li-2@manchester.ac.uk). Tel: +44 (0) 161 529 3084
[§] The corresponding author: Xi Gao (gaoxi1979@126.com).

## Highlights

- An enhanced framework using explainable AI is proposed for property estimation
- The connectivity matrix method is employed to generate molecular features
- Random forest is used for feature pooling while preserving the original features
- Adjusted $R^2$ is used to evaluate the contribution of features to the model
- Shapley value confirms that properties are influenced by molecule structure

## Graphical abstract

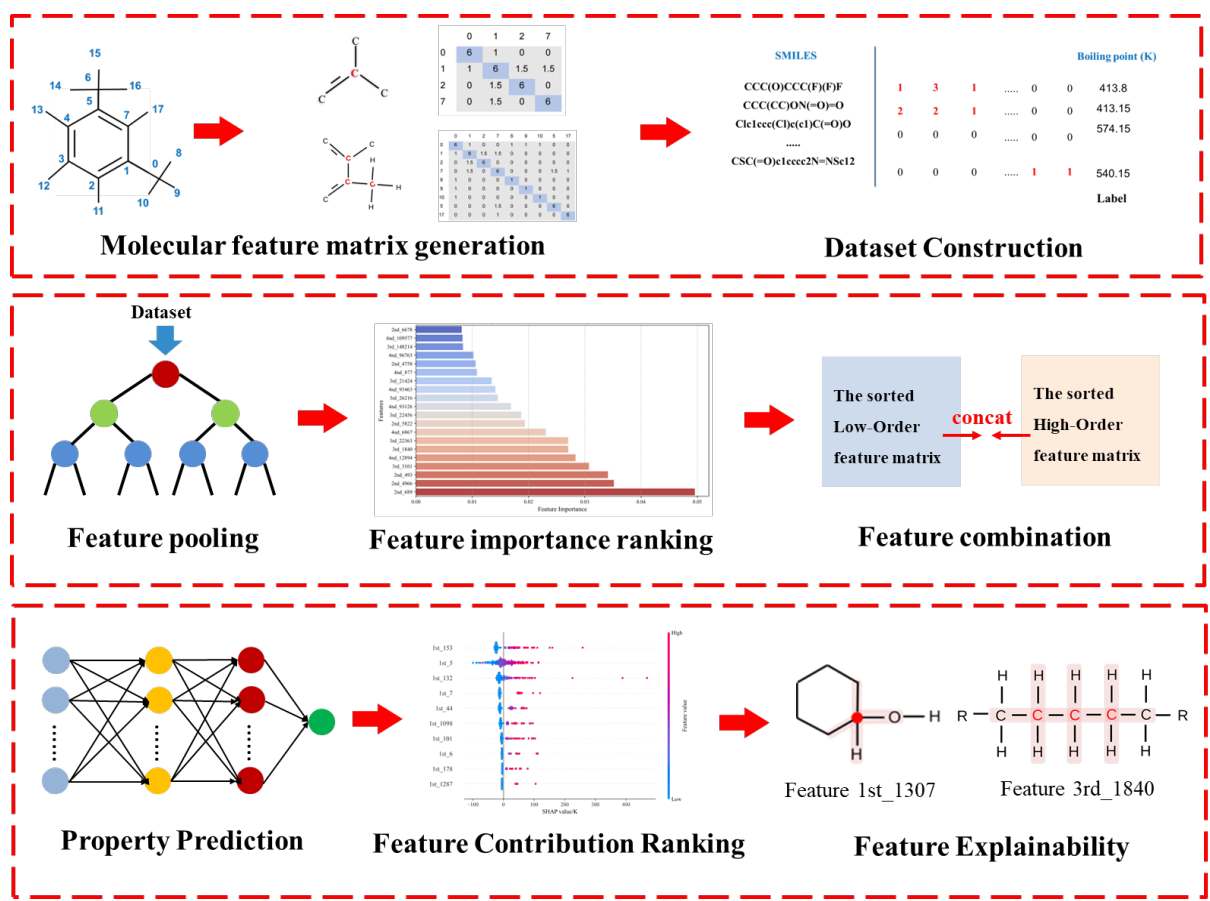

# 1 Introduction

The petroleum refining and chemical industry is the third largest greenhouse gas emitter among all stationary emission sources, accounting for nearly 5% of global greenhouse gas emissions in the energy sector [1]. Growing competitive pressures and increasingly stringent environmental regulations have prompted refineries to enhance resource conversion efficiency and transition toward cleaner production, thereby reducing unnecessary energy and material consumption. Under these circumstances, the concept of "placing the right molecule in the right place at the right time, at the right price"—known as "molecular management"—is highly aligned with the objectives of modern refineries [2,3]. By exercising precise molecular control, molecular management maximizes resource utilization and minimizes ineffective reactions or by-product formation, thus directly supporting the emission reduction goals of clean production [4–6].

With the advancement of computing power and the rapid development of artificial intelligence, computer-aided molecular design (CAMD) has been applied to molecular management [7,8]. Its implementation typically involves three main steps: problem definition, molecular design, screening and selection. Among these, molecular design is the crucial step, initially generating candidate molecules and then predicting their properties. Accordingly, appropriate mathematical models must be developed to estimate a compound's properties based on existing experimental data. Meanwhile, as chemical synthesis technologies continue to evolve, the number of new compounds grow rapidly and their complexity increases, making direct experimental measurements of molecular properties impractical. Hence, developing accurate and efficient methods for estimating pure component properties is an urgent priority.

Currently, group contribution (GC) methods are widely used for estimating pure component properties [9,10]. In this approach, a molecule is viewed as being composed of multiple functional groups. Each group's contribution value and corresponding fitted contribution vectors are then substituted into a linear relationship to estimate the desired property. Gani et al. [11] introduced higher-order groups to improve prediction accuracy. Specifically, first-order groups capture chemical valence and atomic balance, whereas higher-order groups recognize isomers and multi-functional compounds. Building on the GC methods, the phase-equilibrium prediction approach [12] and thermodynamic property prediction method [13] were also subsequently proposed. However, the phenomenon of residual absorption by certain higher-order groups during the fitting process may lead to overfitting, which could reduce the prediction accuracy for unseen molecules [14]. Meanwhile, the proximity effects and nonlinear interaction features between groups remain unrepresented.

The powerful nonlinear representation capability of machine learning helps address some of the aforementioned issues. Alshehri et al. [10] employed gaussian process regression (GPR) on a dataset of over 24000 chemicals, achieving accurate predictions for 25 different properties. Thermodynamic models based on group contribution (GC), such as UNIFAC and SAFT-γ Mie, are commonly employed for predicting thermodynamic properties [15,16]. Cao et al. [17] introduced a warping function based on GPR to encapsulate discrete variables and map them to a continuous domain. Nevertheless, GC is built on a predefined library of functional groups rather than a systematic, automatic feature-representation method, so it cannot guarantee adequate representation of every possible molecule. If a new molecule contains an entirely novel functional group, the prediction becomes unreliable, the results of Cao et al. [17] indicate that the model's performance in predicting the properties of unseen molecules is significantly inferior to its performance on the training set, Aouichaoui et al. [18] suggesting that overfitting may still be present. At the same time, insufficient molecular representation may lead to poor discrimination between isomers, resulting in cases where different molecules share the same feature vectors but have different physicochemical properties. This non-one-to-one mapping between molecular features and properties significantly could reduce the prediction accuracy and generalization ability of the model. In addition, the linear relationships used in conventional GC frameworks cannot capture the nonlinear interactions between different groups, which further contributes to prediction inaccuracies [19]. These shortcomings have motivated researchers to further develop new approaches for molecular structure representation.

Using molecular descriptors to represent molecular structure is another approach to molecular property estimation. Essentially, this method encodes atomic, bonding, and topological indices into descriptors—typically fixed-length numerical vectors—that serve as inputs to machine learning or deep learning models, thereby correlating those vectors with the pure component properties. Visco et al. [20] proposed a Signature Molecular Descriptor that represents molecular structure as a set of hierarchical "signatures" to systematically encode topological information. Rogers et al. [21] developed Extended-Connectivity Fingerprints (ECFPs), assigning an initial integer identifier to each atom and collecting identifiers of neighbouring atoms; these are then hashed into a new, unique identifier. The results showed that ECFPs can accurately and efficiently facilitate property prediction and compound classification. However, because ECFPs rely on hashing, the resultant fingerprint vectors tend to be high-dimensional and sparse, negatively impacting computational efficiency, while bit collisions may cause different molecular structures to map to identical fingerprints. Van Geem et al. [22] established a geometry-based molecular representation that encodes internal distances, bond angles, and dihedral angles, and uses gaussian mixture models (GMM) to learn the statistical distributions of these geometric features, ultimately converting the molecular structure into a fixed-length numerical vector. While these descriptor-based methods can yield accurate predictions, the quality of the results is highly dependent on the descriptors' ability to capture the complexity of chemical structures. Moreover, because the descriptors reduce a molecule's structural features to numerical vectors, their interpretability is relatively weak, making it difficult to draw direct connections between a compound's structure and its properties.

Inspired by methods in natural language processing, Mi et al. [23] proposed an approach to predict the melting points of organic small molecules by parsing SMILES-encoded molecular structures. This method eliminates the need for additional chemical descriptors and demonstrates superior prediction accuracy compared to traditional approaches based on molecular fragments or descriptors. Zeng et al. [24] developed a deep learning system that integrates molecular structures with biomedical textual information, enabling comprehensive cross-source molecular understanding. Goh et al.[25] introduced SMILES2vec, which employs a bidirectional gated recurrent unit and a convolutional neural network (CNN) to extract SMILES-based "linguistic" features without the need for manual feature engineering, thereby using recurrent neural network to predict chemical properties. Subsequently, Mann et al. [26] proposed the Grammar2vec framework, which uses a parse tree to treat SMILES as "sentences" and then leverages the Word2vec model to convert molecular structures into dense vectors. Compared with conventional SMILES representations, Grammar2vec demonstrates advantages from an information-theoretic perspective. Although the aforementioned methods have improved the prediction capability for the properties of unseen molecules, this unsupervised feature extraction approach struggles to establish a clear correspondence between molecular features and specific molecular structures.

Graph neural networks (GNN) have also been used to extract molecular features, representing molecules as graphs in which atoms serve as nodes and bonds as edges [27,28]. Building on this, Ishida et al. [29] incorporated three-dimensional structural features by computing relative coordinate differences, thereby capturing the spatial organization of molecules. These structural features are then combined into molecular fingerprints via convolution and pooling operations, and machine learning models are subsequently employed to correlate these fingerprints with specific properties. Zang et al. [30] proposed a hierarchical molecular graph self-supervised learning framework, which modelled molecular structures using GNN and enhances molecular representation learning through multi-level self-supervised tasks. This method introduces molecular motif nodes and graph-level nodes to achieve hierarchical characterization of molecular structures, while improving model generalizability via multi-tiered prediction tasks. Nevertheless, the convolution operation used in GNN can cause information loss, making it difficult to trace how individual substructures contribute to the overall property profile. In view of this, Aouichaoui et al. [18] constructed feature vectors at three levels including atomic, group, and junction-tree to achieve multi-scale and multi-perspective molecular representations and introduced an attention mechanism to highlight specific functional groups based on the learned attention weights, thereby identifying the most critical features for property prediction. However, the attention mechanism may not always reliably reflect the true causal importance [31], as attention represents only part of the internal weights, and other layers also influence decision-making. Changes in attention may be compensated by other parameters, leaving the outcome unaffected. Many studies have found that even when a component with high attention score is removed, the prediction sometimes does not change significantly [32]. Moreover, the attention mechanism is only applicable to a network with an explicit attention structure, making the model highly dependent on such architecture.

SHAP analysis by calculating the marginal contributions of all feature subsets, it considers the impact of adding or removing features on the model output. In contrast to attention mechanisms, SHAP analysis is grounded in strong axiomatic foundations, and is completely independent of the model structure [33]. Yang et al. [34] developed a hybrid molecular descriptor and predicted the solubility of $CO_2$, calculating Shapley values of individual features to elucidate their importance in the prediction outcome. The results indicated that pressure and temperature were the most influentially operational variables, while molecular characteristics such as bond lengths, bond angles distribution, electronic states, and surface areas were also correlated with solubility. Similarly, Mann et al [26] proposed Grammar2vec to generate dense, numeric molecular representations and perform a Shapley value based analysis to estimate feature importance. However, the specific influence of molecular structures on the prediction results has not been determined in the aforementioned studies, which may be attributed to the inherent limitations of the constructed molecular descriptors in capturing detailed structural information.

To overcome the disadvantages of the aforementioned methods for property estimation of pure compounds, a novel framework for pure component property estimation using explainable machine learning methods is proposed in this work, where a molecule is firstly represented as a connectivity matrix (CM) based on SMILES and then a large number of submatrices from the CM [35] are automatically and systematically extracted. Each submatrix represents the environment of an atom and molecular structure. The frequency of occurrence of these submatrices is transformed into molecular features, which are categorized into low-order and high-order features depending on the number of connected atoms considered during the submatrix construction. Compared to the predefined group contribution-based feature construction, this systematic and automated feature extraction method can enhance the property predictive performance for unseen molecules, meanwhile their proximity effects are thoroughly considered. Subsequently, the supervised learning method, random forest (RF), is utilized for feature pooling, preserving the original semantic meaning of the features. Artificial neural network (ANN) and Gaussian processes algorithms are employed to establish prediction model for estimation of normal boiling point ($T_b$), critical pressure ($P_c$), critical temperature ($T_c$) and liquid molar volume ($L_{mv}$). Adjusted $R^2$ is introduced to evaluate the impact of the number of features on the prediction results, avoiding model overfitting. Finally, Shapley value is employed to systematically understand the importance of the features. The key molecular structures influencing the above properties are thus summarized.

The remainder of this paper is organized as follows. In **Section 2**, the property estimation problem is defined. **Section 3** presents the technical details of the proposed framework. **Section 4** then describes dataset and details of the model training process. In **Section 5**, the results of feature generation and predictive results are presented, and then the most influential features are identified by the Shapley values. **Section 6** concludes the work and discusses potential future directions.

## 2 Problem statement

There are total $I(i=1,2,\ldots,I)$ molecules, A representative molecule may have $P(p=1,2,\ldots,P)$ roperties such as $T_b$, $T_c$, $P_c$ and $L_{mv}$. The properties of each molecule can be obtained from a database or estimated using property prediction models. We regard this task as a regression problem, with the goal of constructing a nonlinear fitting model to establish the mapping between pure component properties and molecular structures, as shown in **Eq. (1)**.

$$\boldsymbol{p}_i = \boldsymbol{f}(\boldsymbol{x}_i, \boldsymbol{\beta}) \quad i \in I \tag{1}$$

where $\boldsymbol{p}_i$ is a vector of thermodynamic properties to be estimated for the $i$-th molecule; $\boldsymbol{x}_i$ is the feature vector of that molecule; $\boldsymbol{\beta}$ is the coefficient vector to be estimated by the regression model; $\boldsymbol{f}(\cdot)$ represents a vector of the nonlinear fitting models.

To achieve accurate property estimation, two key factors are the construction of the feature vector $\boldsymbol{x}_i$ and $\boldsymbol{f}(\cdot)$. In this study, $\boldsymbol{x}_i$ is obtained via the proposed connectivity matrix–based molecular representation method below, and a machine learning algorithm (e.g., ANN) is employed to map these vectors to the target properties.

## 1 Molecular representation and feature generation

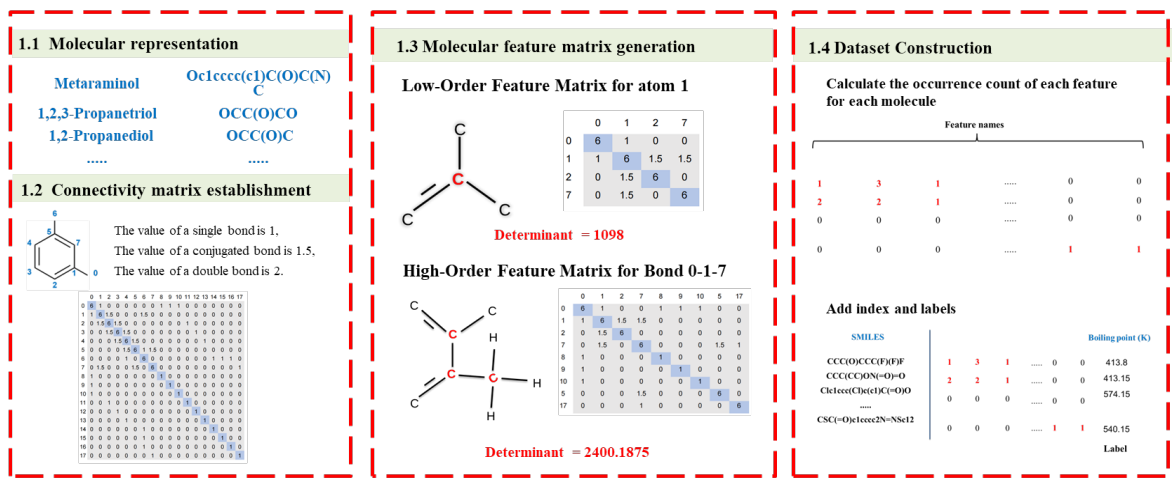

## 2 Feature pooling, ranking and combination

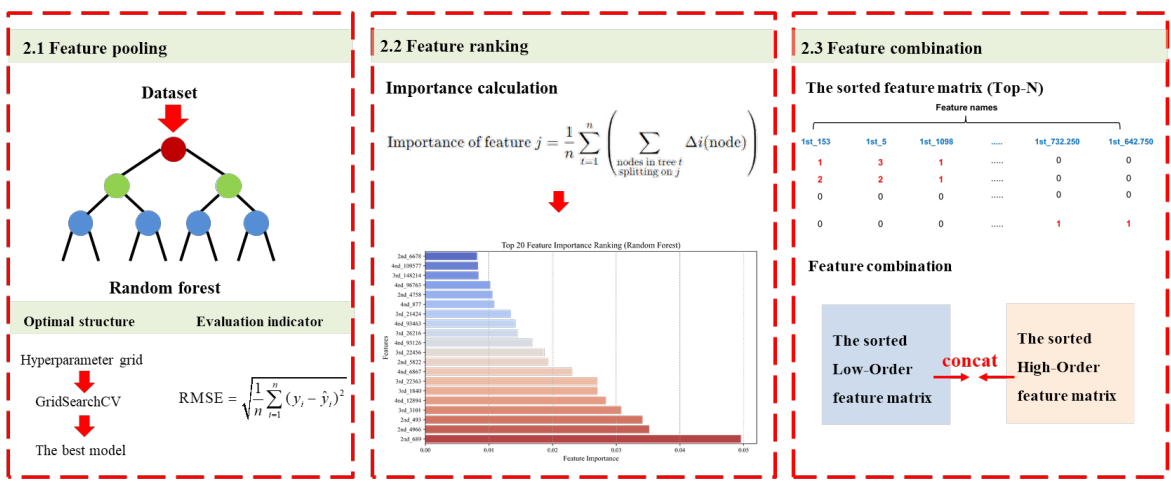

## 3 Property prediction and interpretability analysis

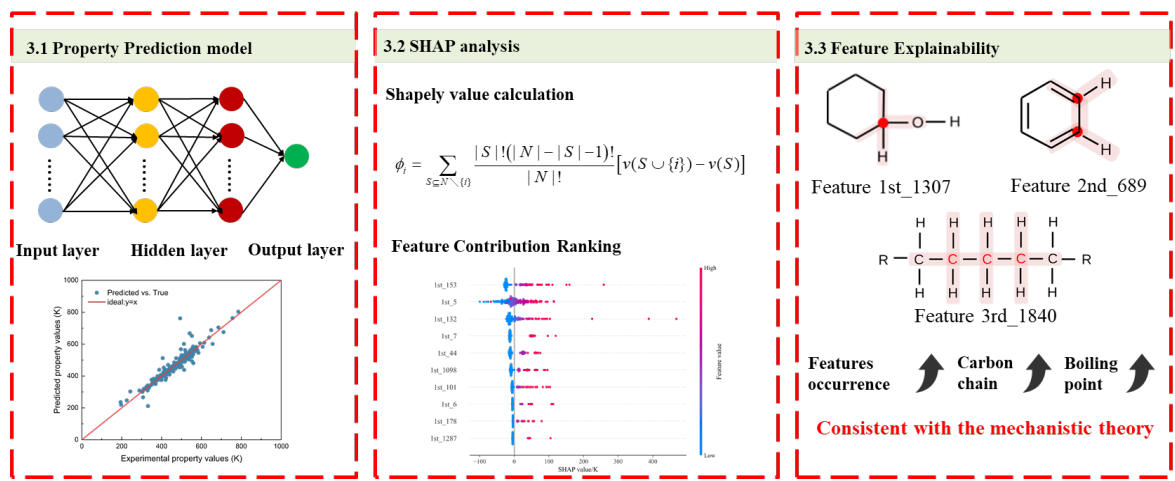

**Fig. 1.** Proposed framework for Pure Component Property Estimation

# 3 Framework for Pure Component Property Estimation

This section provides a detailed introduction to the implementation process of the proposed framework and the methodologies, as shown in **Fig. 1**, the implementation process is as follows: (1) Molecular structures are first represented using the connectivity matrix method based on molecular SMILES. Submatrices for individual atoms are constructed according to their bonding configurations. Depending on different atomic bonding patterns, first-order, second-order, third-order, and fourth-order submatrices can be obtained, these submatrices are further combined to account for the influence of interactions between adjacent atoms on the properties. Furthermore, the frequency of occurrence of these submatrices are regarded as the feature vector of the molecule. (2) The supervised machine learning model RF is applied for feature pooling, and the optimal hyperparameters are determined using the GridSearchCV method [36,37]. To evaluate the impact of the number of retained features on the results, the Adjusted $R^2$ is introduced to penalize the inclusion of additional features, thereby preventing an artificial increase in $R^2$ simply due to feature expansion. (3) A machine learning model (e.g., ANN) is then constructed to predict $T_b$, $T_c$, $P_c$ and $L_{mv}$. (4) Finally, to enhance the interpretability of the model, a feature analysis method based on Shapley values [38,39] is employed to determine the contribution of each feature to the property prediction.

## 3.1 Molecular representation and feature generation

In this study, connectivity matrices are employed to represent molecular structures. Ambiguities in molecular descriptions are eliminated through canonical SMILES representation. For instance, Toluene may appear as either "c1ccccc1C" or "Cc1ccccc1" in original SMILES formats, but its canonical form is standardized as "Cc1ccccc1", ensuring strict uniqueness in atomic indexing. A parsing script is developed using the open-source cheminformatics toolkit RDKit [40] to directly extract atomic coordinates and bond connectivity data from canonical SMILES strings, which are then mapped to adjacency matrices.

In this canonical SMILES-based molecular structure characterization method, all atoms are assigned unique identifiers (IDs) through sequential indexing. To systematically describe molecular topological connectivity, an $N \times N$ adjacency matrix (denoted as $M$) is constructed, where $N$ corresponds to the total number of atoms in the molecular system. The matrix structure is defined as follows: $M = (a_{ij})_{N \times N}$. The diagonal elements $a_{ij}(i=j)$ record the atomic number of corresponding elements, with row/column indices strictly matching atomic IDs. Non-diagonal elements $a_{ij}(i \neq j)$ represent chemical bond types between atoms: elements are assigned 0 when no bond exists between atoms $i$ and $j$, while single, double, aromatic, and triple bonds are represented by values of 1, 2, 1.5, and 3, respectively. This symmetric assignment mechanism ensures complete mathematical characterization of molecular connectivity relationships.

Based on this, the local chemical environment of an atom can be defined by extracting specific submatrices from *M*, where the dimensions are dynamically determined by the neighbourhood radius *k* (i.e., the number of consecutively bonded atomic layers). When $k=1$, the first-order submatrix of an atom is generated by traversing the atoms directly bonded to it (including the atom itself) and extracting the rows and columns in the connectivity matrix *M* whose indices match the ID numbers of these atoms. When $k \geq 2$, the *k* consecutively bonded atoms are treated as a whole, expanding the extraction range to include their direct neighbouring atoms. To prevent overfitting, this study sets $k^{\max}=4$ meaning that through a hierarchical extraction strategy, first, second, third, and fourth-order submatrices can be obtained, providing a scalable foundation for subsequent feature analysis and machine learning modelling. However, even when atoms share identical chemical environments within a molecule, their submatrices may still exhibit variations. Addressing this issue requires constructing an accurate submatrix representation system. Although both the determinant and eigenvalues are fundamental properties of a matrix, matrices derived from different atomic environments may have the same determinant values, rendering determinant-based differentiation ineffective. Notably, eigenvalues demonstrate superior discriminative capability in this context—when an atomic environment changes, its corresponding eigenvalues typically exhibit significant differences. Based on this observation, this study selects eigenvalues as the characterization parameter. Considering the mathematical simplicity of determinant expressions, we first convert determinant values into standard strings as primary identifiers. If duplicate determinant values occur, they are distinguished using suffixes such as "_1", "_2", etc. On this basis, a classification criterion is established: if two submatrices have identical sets of eigenvalues, they are classified as the same category; otherwise, they are assigned to different categories. Through this hierarchical recognition strategy, precise classification and identification of all submatrices are ultimately achieved. Finally, by statistically analysing the number of submatrices with homologous eigenvalues, the characteristic distribution patterns within the molecular system can be effectively represented.

Specifically, let *d* denote the total number of low-order and high-order sub-matrix features for a molecule. The feature profile of a single molecule is then represented by a vector $\boldsymbol{f} = (f_1, \ldots, f_d) \in \mathbb{N}^d$. If an atom has a first-order sub-matrix whose determinant equals 5, and all other first-order submatrices in the molecule with determinant 5 share exactly the same set of eigenvalues, we group them under the common label "1st_5" and record how many times this submatrix appears; that count is the vector element $f_1$. If another first-order submatrix with the same determinant 5 exhibits a different eigenvalue pattern, we separate it into a new category labelled "1st_5_2" and store its frequency as $f_2$. Applying this rule to every order (1st – 4th) and to every unique "determinant + eigenvalues" combination, we obtain the complete sparse feature vector $\boldsymbol{f} = (f_1, \ldots, f_d)$, which serves as the input to the subsequent machine-learning algorithms.

### 3.2 Feature pooling

Due to the diversity of molecular fragments, the generated submatrices are typically sparse, as many entries are zero. Moreover, many features are irrelevant or redundant for regression analysis and exist as noise. The presence of non-informative variables may increase prediction uncertainty, thereby reducing the overall effectiveness of the predictive model. Therefore, feature selection is necessary to reduce data dimensionality and enhance model accuracy. To ensure that all molecules can be described by at least some features, all first-order features are retained, and feature selection is applied only to higher-order feature matrices.

Common feature pooling methods include Principal component analysis (PCA) [41] and Pearson correlation analysis [42]. However, these methods are unsupervised learning techniques that map the original feature vectors into a lower-dimensional space through linear transformations, effectively generating new combined features [41]. This process results in loss of the original feature information, making the final regression model less interpretable and overlooking nonlinear interactions between features. In contrast, the RF method, as a supervised learning-based feature selection strategy, effectively overcomes these limitations [43]. Each decision tree serves as a classifier, and for a given input sample, *N* decision trees will produce *N* classification results. RF aggregates the results of multiple decision trees to form a strong classifier, thereby improving classification performance. The final output is determined by the collective prediction results of all decision trees.

For a regression problem, the first step is to calculate the average decrease in impurity (variance) caused by each feature when splitting at all tree nodes, as shown in **Eqs. (2)-(3).**

$$Var(D) = \sum_{j \in D} (y_j - \overline{y}_D)^2 / |D| \tag{2}$$

$$\overline{y}_D = \sum_{j \in D} y_j / |D| \tag{3}$$

where $D$ is a set of molecules input to a node, which should be a subset of set $I$; $y_j$ is measured (or target) value of molecules $j$, and $\overline{y}_D$ represents the mean of all measured (target) values.

The total impurity (variance) reduction contributed by a specific feature within a single tree is computed to determine its feature importance, as shown in **Eq. (4)**:

$$\Delta Var = Var(D) - \left[ \frac{N_L}{N} Var(D_L) + \frac{N_R}{N} Var(D_R) \right] \tag{4}$$

Where $D_L$, $D_R$ and $D$ represent the set of samples (molecules) at the left child, right child and the parent nodes, respectively. $N$ is the number of samples (molecules) in the parent node, while $N_L$ and $N_R$ are the numbers of samples (molecules) at the left and right child nodes, respectively.

The feature importance across all trees is calculated, as shown in **Eqs. (5) – (6)**. Finally, the features are ranked, and the top $n$ features with the highest contribution are selected.

$$Importance \cdot of \cdot feature_j^{(single\ tree)} = \sum_{all \cdot nodes\ where \cdot X_j \cdot is \cdot used} \Delta Impurity_j \tag{5}$$

$$Importance \cdot of \cdot feature_j = \sum_{t=1}^{T} Importance \cdot of \cdot feature_j^{(t)} / T \tag{6}$$

The above method is known as the Top-N algorithm [44], which directly selects original features rather than generating linear combinations. Additionally, based on the splitting rules of the tree model, it can automatically identify complex nonlinear relationships between features and the target variable. To ensure the correct ranking of feature importance, GridSearchCV [36] is used to determine the optimal hyperparameter combination for the RF model.

For regression problems, the $R^2$ value of the test set is often used as an evaluation metric to reflect the model's explanatory performance and generalization ability. A higher $R^2$ value indicates that the model explains a larger proportion of variance, as shown in **Eq. (7)**. However, $R^2$ has a notable drawback: when irrelevant variables are added to the model, $R^2$ artificially increases, even if these variables do not actually improve the model's predictive ability. This can lead to overfitting, especially in cases where the number of variables is large.

$$R^2 = 1 - SS_{res} / SS_{tot} \tag{7}$$

where $SS_{res}$ is the residual sum of squares, $SS_{res} = \sum_{i=1}^{n}(y_i - \hat{y}_i)^2$ and $SS_{tot}$ is the total sum of squares, $SS_{tot} = \sum_{i=1}^{n}(y_i - \overline{y})^2$.

To mitigate this issue, the adjusted $R^2$ is introduced, as shown in **Eq. (8)**. Compared to the standard $R^2$, the adjusted $R^2$ incorporates a degree-of-freedom correction factor $\frac{n-1}{n-k-1}$ to penalize model complexity (i.e., the number of independent variables, $k$). When a new feature is introduced, if it does not improve the model's fitting performance (i.e., $R^2$ remains unchanged), the degree-of-freedom correction factor decreases, leading to a reduction in adjusted $R^2$. By evaluating the difference between adjusted $R^2$ and $R^2$, the risk of overfitting due to the inclusion of irrelevant variables can be effectively avoided.

$$\text{adjusted } R^2 = 1 - \left[(1-R^2)(n-1)/(n-k-1)\right] \tag{8}$$

where $n$ is the number of samples, and $k$ is the number of features.

Finally, the RF feature importance ranking method is used to independently evaluate the importance of first-order features $F_1^{(M)}$ and higher-order features $F_2^{(N)}$, sorting them separately. Based on this, starting from $M_0 = 20$ and $N_0 = 20$, features are incrementally added in a step of 20, following a descending order of importance, to construct dynamically increasing feature subset sequences $F_1^{(M)}(M = 20, 40, ..., M_{max})$ and $F_2^{(N)}(N = 20, 40, ..., N_{max})$. These subsets are then combined to form the joint feature space $F_{final}^{(M,N)} = F_1^{(M)} \cup F_2^{(N)}$. By training a property prediction model and calculating adjusted $R^2$, the optimal number of features is determined by maximizing adjusted $R^2$.

### 3.3 Property Prediction model

Regression analysis serves as a fundamental method for establishing the mapping relationship between independent and dependent variables, For regression tasks, both parametric and non-parametric models can be employed. To validate the generality of the CM-based molecular representation proposed in this study, it is evaluated using both a classical parametric model, ANN [35,45], and a non-parametric model, GPR [46], both of which have been previously used for property prediction.

### 3.3.1 ANN model

The present study employs an ANN architecture to construct a molecular property prediction model, with the design as follows: molecular feature vectors serve as the input layer, while target property parameters form the output layer. The relationship between features and properties is established through nonlinear transformations in the hidden layers. The model is developed using Python Keras [47] and tensorflow [48] frameworks, implementing a typical feedforward neural network structure, which consists of an input layer, hidden layers, and an output layer. For a neuron $j$ in layer $i$ (denoted as $j_i$), its input and output are calculated by **Eqs. (9) – (10)**.

$$x^i_{j_i} = \sum_{j_{i-1}=1}^{q_{i-1}} w^{i-1}_{j_{i-1},j_i} \cdot x^{i-1}_{j_{i-1}} \tag{9}$$

$$y^i_{j_i} = f(x^i_{j_i} + b^i_{j_1}) \tag{10}$$

The mean squared error (MSE) is adopted as the loss function to measure the discrepancy between the predicted and actual values in regression tasks, where a smaller value indicates higher prediction accuracy, as shown in **Eq. (11)**.

$$\mathcal{L} = \frac{1}{N}\sum_{n=1}^{N} \| y_n - \hat{y}_n \|^2 \tag{11}$$

where $y_n$ is the experimental property value of the *n*-th sample; $\hat{y}_n$ represents the predicted value of the network, and $N$ denotes the total number of samples.

L2 regularization is employed to suppress excessive weights and mitigate overfitting by penalizing large weight values and smoothing the weight distribution, accordingly, the new loss function can be expressed as **Eq. (12).**

$$\mathcal{L}_{reg} = \mathcal{L} + \frac{\lambda}{2}\sum_{l=1}^{L} \left\| W^{(l)} \right\|^2 \tag{12}$$

where $W^{(l)}$ represents the weights of all neurons in layer *l*, and $\lambda$ is the regularization coefficient that controls the strength of the penalty on the magnitude of the weights.

A 6-fold cross-validation strategy [49] is employed to fully explore the statistical properties of the dataset and assess the impact of feature perturbations on the model's generalization ability. Specifically, the original dataset is evenly divided into six mutually exclusive subsets, with each iteration selecting five subsets as the training set and the remaining one subset as the validation set. To enhance computational efficiency and parameter optimization stability, the Adam optimizer is introduced. During backpropagation, network weights and biases are iteratively updated by minimizing the loss function. In addition to $R^2$ and adjusted $R^2$, a multi-metric validation strategy is applied to each fold to avoid evaluation bias caused by differences in metric scales, incorporating RMSE [50], Mean Absolute Error (MAE), and Mean Absolute Percentage Error (MAPE) [51], as shown in **Eqs. (13) – (15)**.

$$\text{RMSE} = \sqrt{\frac{1}{N}\sum_{n=1}^{N}(y_n - \hat{y}_n)^2} \tag{13}$$

$$\text{MAE} = \frac{1}{N}\sum_{n=1}^{N} | y_n - \hat{y}_n | \tag{14}$$

$$\text{MAPE} = \frac{100\%}{N}\sum_{n=1}^{N} \left| \frac{y_n - \hat{y}_n}{y_n} \right| \tag{15}$$

### 3.3.2 GPR model

The performance of the GPR model is exclusively dependent on the fitness of its covariance (or kernel) function. The goal of the covariance function is to control the shape and characteristics of the fitted function by defining the correlation between the *i*-th and *j*-th molecules, given by **Eqs. (16) – (19)**.

$$y_{pred} = K_{1\times M} W_{M\times 1} \tag{16}$$

$$W = (K_{M\times M}^{-1}) y_{train} \tag{17}$$

$$Cov_{1\times 1} = K_{1\times 1} - K_{1\times M} U_{M\times M} K_{M\times 1} \tag{18}$$

$$U_{M\times M} = (K_{M\times M} + \sigma^2 I)^{-1} \tag{19}$$

where $y_{pred}$ is the property value that needs to be predicted for the test samples; $K$ is the covariance matrix, which is constructed using the kernel function $k$; $K_{1\times M}$ is the covariance matrix between the test samples and the $M$ training samples; $K_{M\times M}$ is the covariance matrix of the $M$ training samples; $y_{train}$ is the target value vector corresponding to the training set; Cov is the uncertainty; $\sigma^2$ is the observation noise variance.

The radial basis function (RBF) is selected as the covariance kernel, which indicates the degree of correlation between the two variables, as shown in **Eq. (20)**. For a fair comparison, we adopt a kernel function structure consistent with that used by Alshehri et al. [10], which consists of the sum of four RBF kernels. The model is implemented using the scikit-learn module in Python [52]. The two parameters of the kernel function, the length scale $\ell$ and the signal variance $\sigma$, control the horizontal and vertical variations of the function, respectively.

$$k(x_i - x_j) = \sum_{i=4}^{4} \exp\left(-\frac{1}{2\ell_i^2} \|x_i - x_j\|^2\right), k \in K \tag{20}$$

where $x_i$ and $x_j$ represent the feature vectors of two molecular samples; $\|x_i - x_j\|$ is the Euclidean distance between them.

### 3.4 SHAP analysis

To enhance the interpretability of the ANN and GPR models and identify the molecular structures that contribute most to the property prediction, Shapley values are used to quantify each feature's contribution to the prediction results, providing insights into feature importance in the estimation task [53]. This method has been widely adopted in machine learning model interpretability analysis [38,54].

Shapley values measure the average marginal contribution of a feature across all possible feature combinations. Given a model with *n* features, the Shapley value considers all possible feature subsets and their marginal contributions in different permutations. Specifically, it enumerates all non-empty subsets $S \subseteq F \setminus \{i\}$ of the feature space $F_1^{(M)}$ and $F_2^{(N)}$, computing the difference in model prediction before and after adding feature *i* to subset *S* (i.e., its marginal contribution). The contribution of feature *i* to the final prediction is calculated using **Eq. (21).**

$$\phi_i = \sum_{S \subseteq F \setminus \{i\}} \frac{|S|!(n-|S|-1)!}{n!} \left[ f(S \cup \{i\}) - f(S) \right] \tag{21}$$

where $F \setminus \{i\}$ represents the set of all features except feature *i*; $|S|!$ represents the ordering of the features in subset S; $(n-|S|-1)!$ represents the ordering of the remaining $(n-|S|-1)$ features that are not included in subset *S* and do not include feature *i*; $f(S \cup \{i\}) - f(S)$ represents the marginal contribution of feature *i*; $\frac{|S|!(n-|S|-1)!}{n!}$ to ensure that the probabilities of all subset combinations are balanced.

## 4 Dataset ·description and model training

This section presents the details of the database used, along with the hyperparameter selection for the RF model and the ANN model. Section 4.1 describes the dataset used in this paper and employs joint conditional entropy to demonstrate the superiority of the connectivity matrix-based feature construction method over the group contribution method. Section 4.2 presents the division strategy for the training, validation, and test sets. Section 4.3 details the hyperparameter selection for the RF model and the ANN model.

### 4.1 Dataset description

The database used in this study is based on the dataset published by Alshehri et al. [10], which includes 25 pure component properties and 424 different functional groups for each sample, with molecular samples represented using SMILES strings.

This work selects $T_b$, $T_c$, $P_c$ and $L_{mv}$ as the target properties due to their significance in characterizing thermodynamic phase behaviour and their relevance in industrial applications. $T_b$, as a key indicator of molecular volatility, not only determines the operating conditions and energy efficiency of separation and purification processes [55]. The ratio of $T_c$ to $P_c$ is often used in the calculation of many equations of state, such as the Peng–Robinson equation of state [56]. Therefore, accurate knowledge of $T_c$ and $P_c$ is essential for reliable prediction of Pressure–Volume–Temperature (PVT) relationships. Accurate prediction of $L_{mv}$ provides a foundation for reactor volume design, optimization of mass and heat transfer, and indirect estimation of liquid transport properties such as density, viscosity, and diffusion coefficient [57].

Due to the accessibility and experimental cost associated with different molecular properties, the number of samples varies across properties. As a fundamental indicator of volatility, $T_b$ has the highest data coverage in the dataset (i.e., 5276 samples), as its experimental determination can be conducted under mild conditions (atmospheric pressure and standard temperature) and follows well-established standardized testing protocols (e.g., ASTM D86). In contrast, the measurement of $T_c$ and $P_c$ requires precise control of high temperature and pressure near the critical point, making the experiments more costly and posing decomposition risks for thermally sensitive compounds (such as polycyclic aromatic hydrocarbons and energetic materials); thus, the number of available samples is lower, with 777 and 775 samples for $T_c$ and $P_c$, respectively. The acquisition of $L_{mv}$ relies on high-precision density measurements, and its data scarcity is directly related to experimental cost, complexity in phase control, and stringent requirements for validating consistency across data sources. As a result, the dataset size for $L_{mv}$ is significantly smaller than that for ambient-pressure properties, with a total of 1057 samples.

The distributions and kernel density estimation (KDE) curves for $T_b$, $T_c$, $P_c$ and $L_{mv}$ are shown in **Fig. 2**. The KDE curve is a non-parametric method for estimating the probability density distribution of data, effectively capturing the overall distribution trend [58].

The distribution of $T_b$ values is primarily concentrated in the range of approximately 400–600 K, exhibiting a distinct unimodal characteristic, which indicates a high data density within this interval. This distribution suggests that most samples have relatively centralized $T_b$ values, while a small number of samples exhibit long-tail distribution at higher temperatures (above 1000 K) and near room temperature. For $T_c$, the values are primarily concentrated in the range of 500 K to 700 K, with a peak around 600 K. The KDE curve indicates that the overall distribution is relatively symmetrical, with low skewness. For $P_c$, the histogram shows that most values fall within the range of 20 bar to 40 bar, with the highest frequency near 25 bar. As observed from the KDE curve, the distribution exhibits a long tail extending into the high-pressure region. In contrast, $L_{mv}$ values are mostly concentrated within a narrow range (approximately 0.1–0.2 cc·mol$^{-1}$) and follow a unimodal distribution. For property prediction, the high concentration of data around the peak implies that the main portion of the dataset aligns well with conventional statistical model assumptions, facilitating the ability of machine learning algorithms to accurately learn feature patterns in this range. The presence of extreme values presents both challenges and opportunities for model generalization over a broader range. Overall, this dataset provides strong representativeness and applicability in property prediction, serving as a viable foundation for machine learning model development and validation.

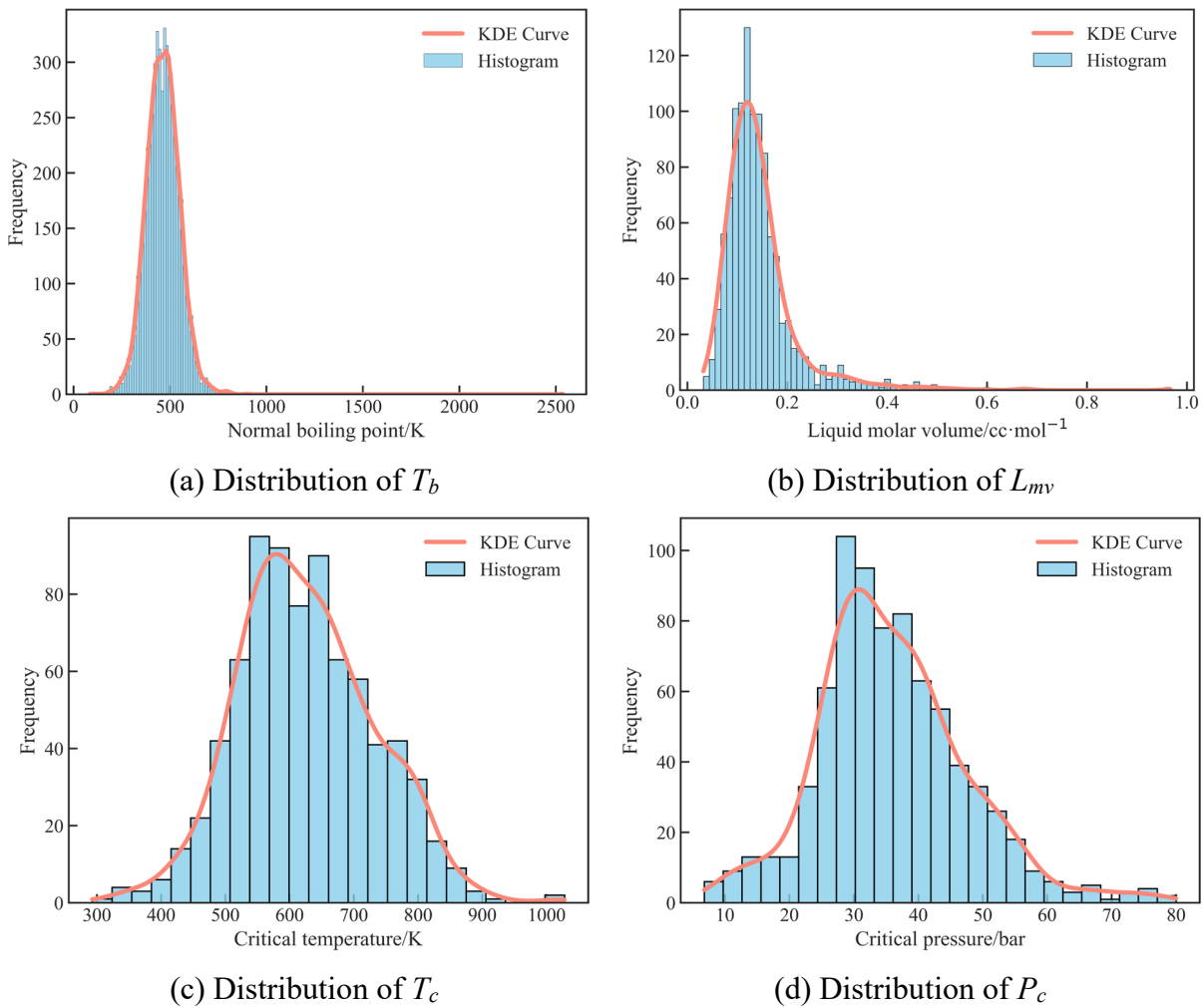

(a) Distribution of $T_b$

(b) Distribution of $L_{mv}$

(c) Distribution of $T_c$

(d) Distribution of $P_c$

**Fig. 2.** Data distribution and kernel density estimation

The molecular sample features in the original dataset were constructed using the GC method, with each molecule represented by a 424-dimensional sparse vector [10]. We refer to this as the GC-dataset. To further characterize the diversity within the GC – dataset, we first count the number of molecules that share the same label but have different feature vectors when using the group contribution method for molecular feature construction. We then perform the same analysis on the dataset constructed using the CM-based approach proposed in this study (referred to as the CM-dataset). The results are summarized in **Table 1**. It can be observed that, under the same feature dimensionality, the connectivity matrix-based feature construction method demonstrates a superior ability to distinguish between molecules compared to the traditional group contribution method.

When all high-order features are included, the total feature dimensionality reaches 12897. Furthermore, we introduce the joint conditional entropy to evaluate the uncertainty of the target property given the feature set, as defined in **Eq. (20)** [59]. A lower value of joint conditional entropy indicates a stronger determinative relationship between features and labels, meaning reduced uncertainty in the predicted property. The theoretical minimum value is zero.

$$H(Y/X) = \sum_x p(x) H(Y/X = x) \tag{20}$$

where $H(Y/X)$ represents the entropy of $Y$ given a specific value of $X$, which quantifies the uncertainty of $Y$ under that condition, bit; $p(x)$ denotes the probability that the random variable $X$ takes the value $x$.

**Table 1** Impact of Different Feature Construction Methods on Joint Conditional Entropy

| Dataset | Feature dimensionality | No. of molecules* | | | | $H(Y/X)$/bit | | | |
|---|---|---|---|---|---|---|---|---|---|
| | | $T_b$ | $T_c$ | $P_c$ | $L_{mv}$ | $T_b$ | $T_c$ | $P_c$ | $L_{mv}$ |
| GC-dataset | 424 | 748 | 67 | 66 | 69 | 0.306 | 0.177 | 0.116 | 0.148 |
| CM-dataset | 424 | 595 | 54 | 54 | 53 | 0.233 | 0.0931 | 0.047 | 0.105 |
| CM-dataset | 12897 | 200 | 20 | 20 | 25 | 0.066 | 0.0077 | 0.0051 | 0.047 |

*Molecules with different features but identical property values

The results in **Table 1** indicate that, under the same feature dimensionality, the CM-dataset exhibits lower joint conditional entropy. This suggests that the mapping between features and labels is more distinct, making it easier for the model to learn the underlying relationship and achieve better fitting performance. Given the input features, the output is almost deterministic, which further implies that the CM-dataset has low noise and low heterogeneity.

**4.2 Training, validation and test subset partition**

For this prediction task, 95% of the data is used for training and validation, while the remaining 5% is reserved for testing. The training and validation data are further divided into six folds for cross-validation [35] as illustrated in **Fig. S1** of the **Supplementary Materials**. At each iteration, a fold is selected as the validation set to monitor the training process, while the remaining five folds are combined to form the training set. The test set is used exclusively to evaluate the model's final performance. This strategy ensures that each sample participates in validation exactly once, allowing for a comprehensive assessment of the model's robustness across different data subsets. The test set remains frozen throughout the model development process and is only used for final performance evaluation after hyperparameter optimization is completed. This design strictly adheres to the "training-validation-testing" three-stage separation principle, ensuring the objectivity of model performance metrics.

**4.3 Model Hyperparameter Selection**

As previously mentioned, to ensure the accurate computation of feature importance,

GridSearchCV, based on the Scikit-learn framework [52], is used to determine the optimal hyperparameter combination for the RF model. By systematically searching through a predefined parameter space and applying a five-fold cross-validation strategy, the best-performing hyperparameter configuration, which maximizes the model's generalization performance, is obtained as follows: *max_depth* = 30, *min_samples_leaf* = 2, *min_samples_split* = 2, and *n_estimators* = 500. A deeper tree structure enables the model to capture high-order nonlinear feature interactions, while imposing constraints on leaf node sample size and split conditions helps mitigate the risk of overfitting. Additionally, the ensemble of 500 decision trees further improves model stability by reducing variance.

For ANN, hyperparameter settings are determined based on preliminary experiments and domain expertise. The *learning rate* is set to 0.0001 to prevent oscillations during the gradient update process. The number of neurons in the hidden layer is 500, and the *EarlyStopping* parameter is set to 100, meaning that training is terminated if the validation loss does not improve for 100 consecutive epochs. This strategy effectively prevents overfitting by monitoring the convergence of the validation loss curve and stopping training when no further improvement is observed [60].

For GPR, in order to compare with the results of Alshehri et al. [10], we adopt the same training approach as theirs. Specifically, 95% of the data is used as training with 20-fold cross-validation. The length scale, signal variance, and noise variance are iteratively optimized by maximizing the log marginal likelihood. Similarly, the initial values of the kernel's scale parameters and hyperparameters are kept consistent with theirs, with the four initial length scale values set to 0.5, 1, 2, and 5, the noise parameter $\alpha$ set to $1\times10^{-5}$, *n_restart_optimizer* set to 5, and *random_state* fixed at 19.

## 5 Results and discussion

This section presents the detailed results obtained by using the proposed framework. Section 5.1 uses meta-xylene as an example to illustrate generation of the first-order and higher-order feature matrices and the final feature vector. Section 5.2 provides the prediction results and analysis of $T_b$, $T_c$, $P_c$ and $L_{mv}$. Section 5.3 conducts an interpretability analysis of the developed ANN model and identifies the molecular structural features with the greatest impact on the prediction results, Finally, we provide the correspondence between common molecular structures and their features for reference.

### 5.1 Results of submatrix feature generation

We use meta-xylene as an example to illustrate the framework. First, the proposed framework is used to process the SMILES encoding with the RDKit cheminformatics toolkit, obtaining the standard SMILES code for meta-xylene as 'Cc1cccc(C)c1'. Each atom is then assigned a unique identifier, and finally the molecular connectivity matrix is constructed according to the method described in Section 3.1, as shown in **Fig. 3**.

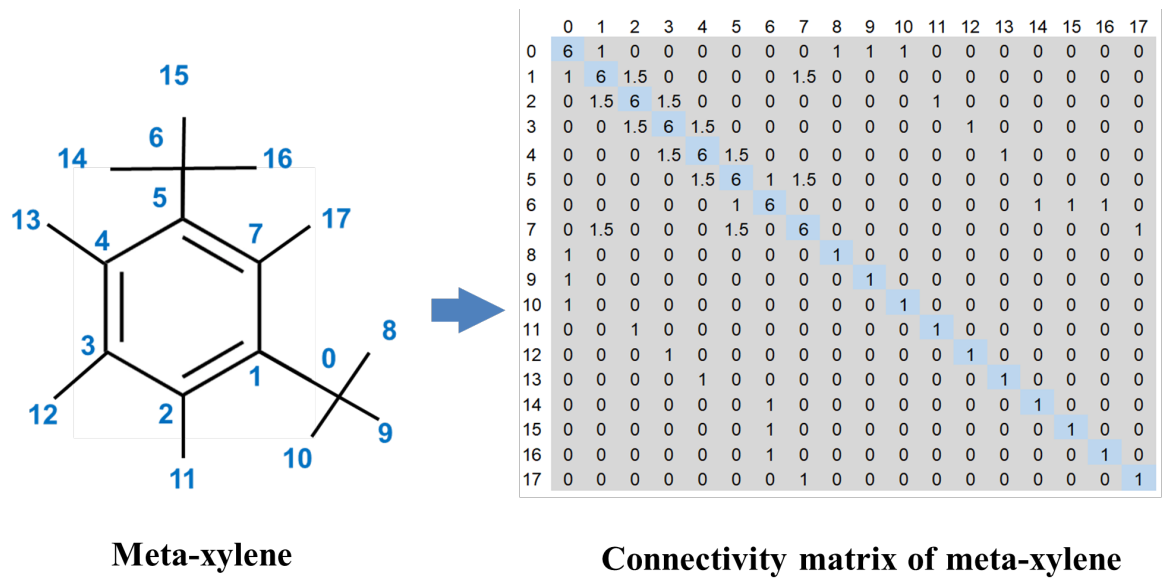

Meta-xylene      Connectivity matrix of meta-xylene

**Fig. 3.** Connectivity matrix of meta-xylene.

Taking atom 1 in meta-xylene as an example, its first-order and higher-order features are derived based on its connectivity relationships, as shown in **Fig. 4**. For the first-order features, the determinant of its first-order environment matrix is calculated as 1098, and it is labelled as "1st_1098." Treating atoms 0 and 1 as a whole, the determinant of their connectivity matrix is computed, and the feature is labelled as "2nd_531." Similarly, considering atoms 0, 1, and 7 as a whole, the determinant of their connectivity matrix is calculated, resulting in the feature label "3rd_17258.625."

Subsequently, all atoms in the molecule are traversed to compute their respective order features. The frequency of occurrence of these features are recorded and the molecular feature vector is then constructed. By applying the proposed framework to the SMILES codes of all molecules in the dataset, the corresponding feature vectors are obtained, forming the overall feature matrix for the dataset. The first-order feature set contains 419 features, while the higher-order feature matrix consists of 12897 features. **Fig. 5** illustrates a schematic of the first-order feature matrix. For $T_b$ prediction, there are 5276 samples, resulting in a first-order feature matrix with dimensions of 5276 × 419.

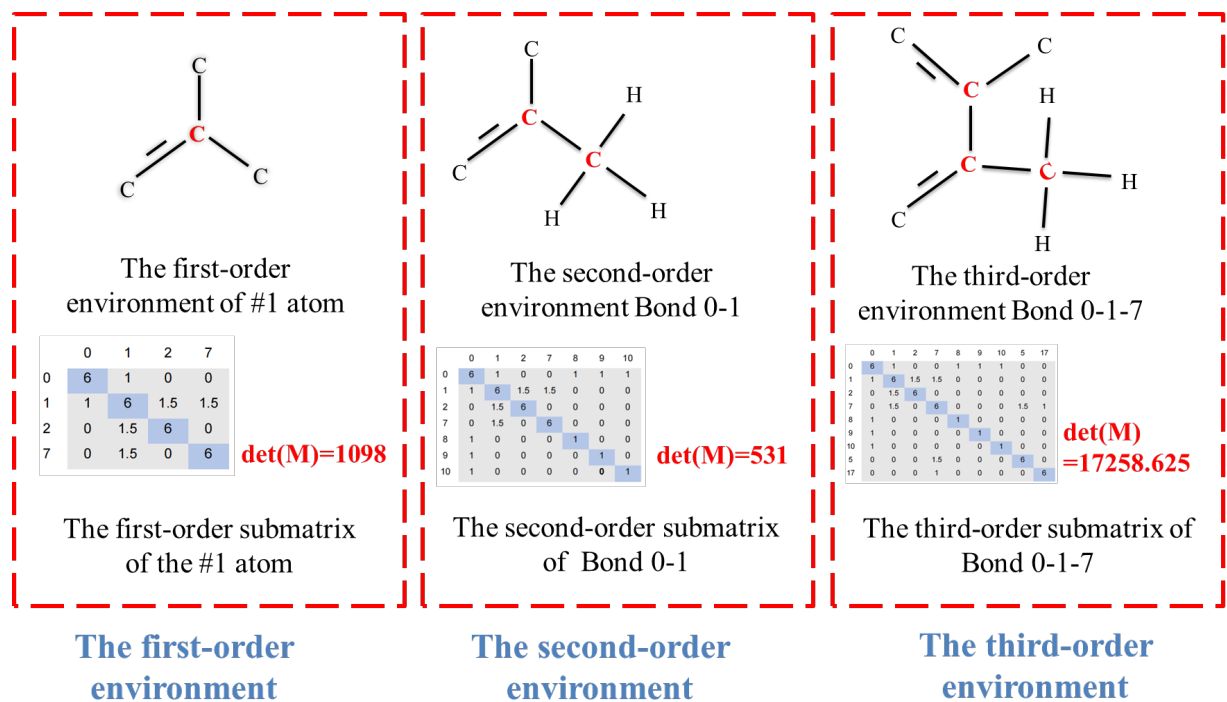

**Fig. 4.** First-order and higher-order features of atom 1 in meta-xylene

**Fig. 5.** First-order features for some molecules in the dataset

**Figs. 6–7** present the feature importance results of the first-order and higher-order feature matrices based on the RF model for $T_b$ prediction. The feature contributions for both matrices are quantified separately where only the top 20 features are displayed. Following the method described in Section 3.2, these features are ranked by importance, and an incremental feature selection approach is applied. The feature subset is gradually expanded in steps of 20 to evaluate the impact of feature space dimensionality on prediction performance.

From the computational results in **Figs. 6-7**, it is evident that first-order features are more important than the higher-order features in $T_b$ prediction. The importance of feature 1st_153 is 0.18, whereas the most important higher-order feature, 2nd_689, has an importance of 0.048, meaning that 1st_153 is 3.75 times more influential than 2nd_689. Additionally, the fifth-ranked first-order feature, 1st_1287, has an importance comparable to 2nd_689. Among the top 20 first-order and higher-order features, the number of features with an importance greater than 0.025 is seven in both cases. However, considering the total number of first-order and higher-order features, which are 419 and 12897, respectively, it can be concluded that the first-order features play a more critical role in the prediction of $T_b$.

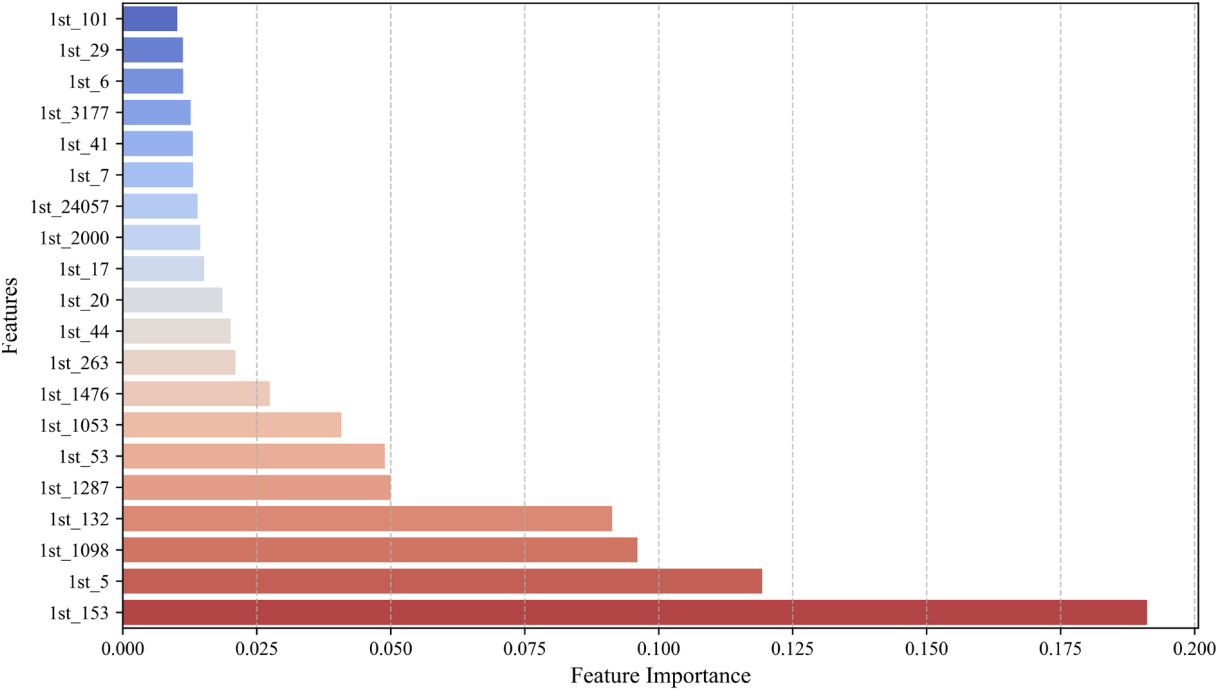

**Fig. 6.** Feature importance results for the Top 20 first-order features in $T_b$ prediction

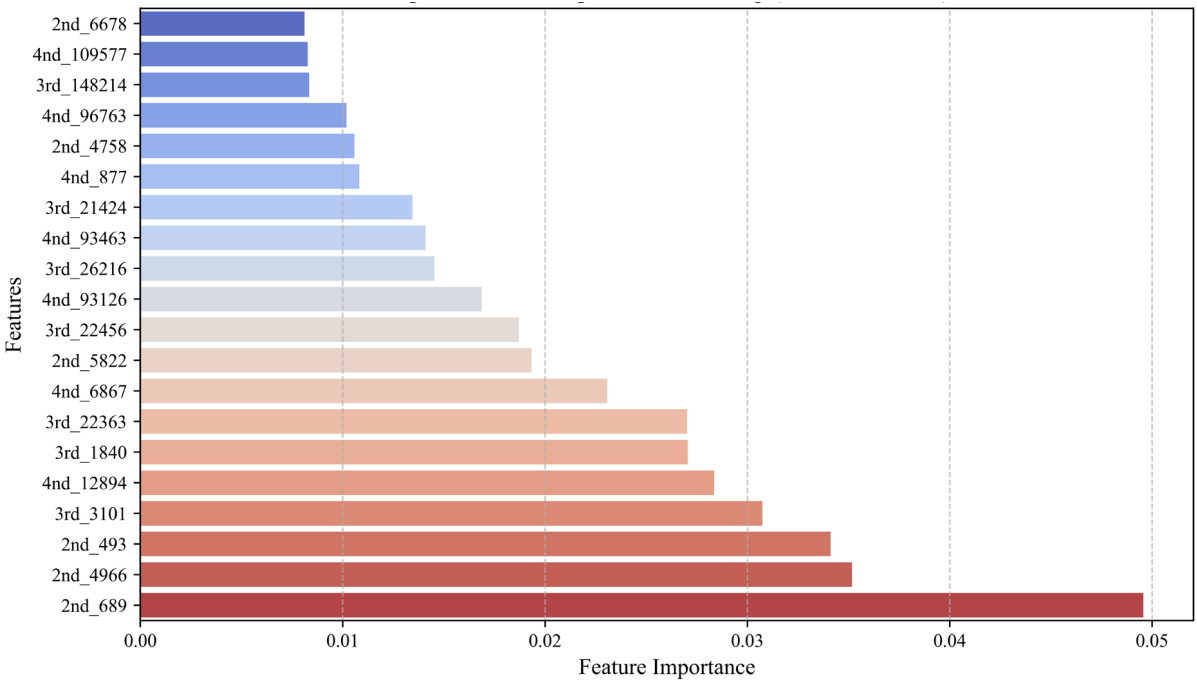

**Fig. 7.** Feature importance results for the higher-order features in $T_b$ prediction

## 5.2 Prediction results

Taking $T_b$ prediction as an example, **Fig. 8** illustrates the performance of $R^2$ and RMSE on the test set as the feature subset expands. Notably, for each feature subset size, six-fold cross-validation is performed, meaning that a total of 126 predictions is executed considering different feature counts and training sets. The final $R^2$ and RMSE values for each feature count correspond to the best-performing ANN model among the six folds. The results show that when the number of features reaches 160, $R^2$ attains its maximum value of 0.917, and RMSE reaches its minimum value of 24.07, indicating the best model performance at this feature dimension.

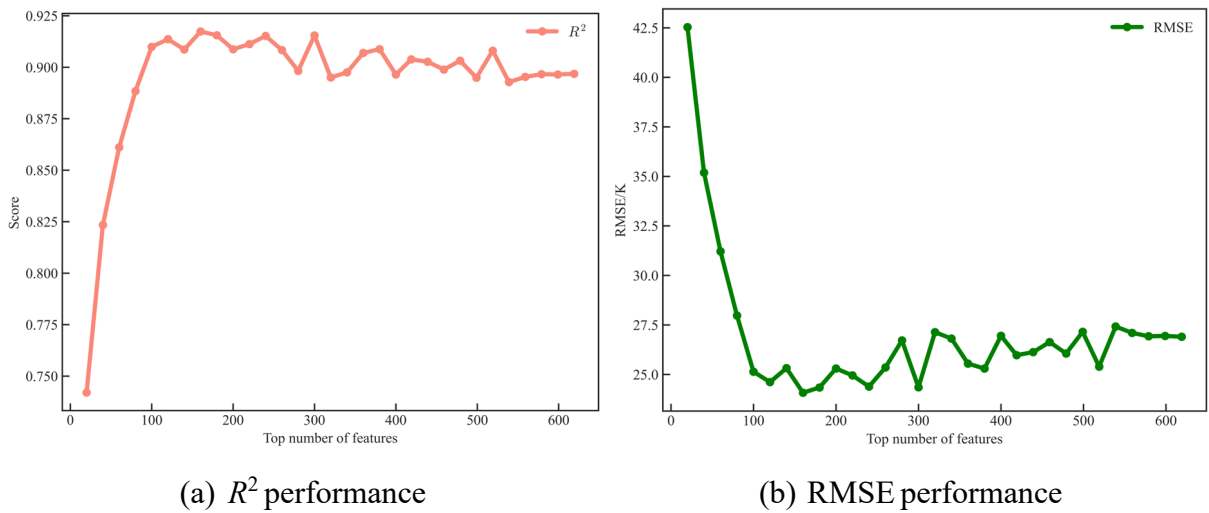

(a) $R^2$ performance    (b) RMSE performance

**Fig. 8.** Performance of the ANN model with increase of feature numbers for $T_b$ prediction

From **Fig. 8 (a)**, it is evident that when the number of features increases from 1 to 100, $R^2$ rapidly rises from an initial value of 0.742 to 0.910. Beyond this point, $R^2$ oscillates around 0.9, indicating that the additional features contribute minimally to the model performance. This observation aligns well with the feature importance results from the RF model, where the importance of the 254th feature is zero, as shown in **Fig. S2** of the **Supplementary Materials**. This suggests that subsequent features do not provide meaningful contributions to the model's predictive capability.

Furthermore, adjusted $R^2$ is computed, with the results shown in **Fig. 9.** It can be observed that when the number of features reaches 100, adjusted $R^2$ attains its maximum value of 0.85. Although the introduction of additional features slightly increases adjusted $R^2$, the improvement is significantly smaller than the penalty introduced by the increasing number of features, leading to a decline in adjusted $R^2$. This indicates that the newly added features contribute minimally to the model's performance, which is consistent with our previous conclusions.

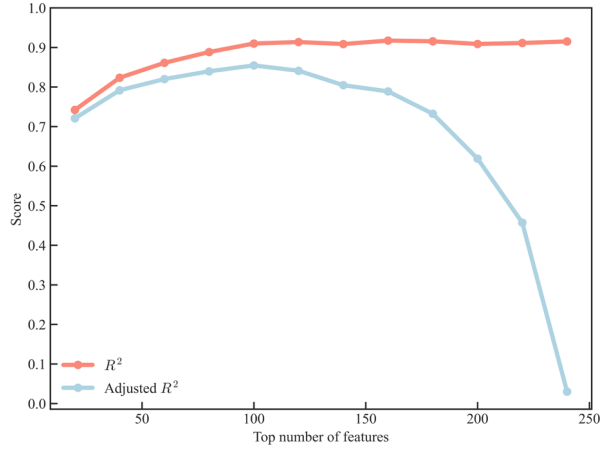

**Fig. 9.** Result of $R^2$ and adjusted $R^2$ of the ANN for $T_b$ prediction with the increase in the number of features

**Table 2** presents the six-fold cross-validation results when the number of selected features in the Top-N approach is 160. In each fold, one subset is used as the test set, another as the validation set, and the remaining four subsets are used for training. Since each fold utilizes different data subsets, variations in training error (RMSE and MAPE) as well as validation and test errors (RMSE and MAPE) are inevitable. Additionally, ANN are subject to random initialization and hyperparameter search variability across different folds. Even under the same hyperparameter settings, slight differences in convergence behavior may arise due to the randomness in initial weights.

**Table 2** Six-fold cross-validation results for $T_b$ prediction (feature number = 160)

| Fold | RMSE/K | | | MAPE | | |
|---|---|---|---|---|---|---|
| | Training | Validation | Test | Training | Validation | Test |
| 1 | 36.53 | 36.26 | **29.33** | 0.038 | 0.039 | 0.041 |
| 2 | 33.17 | 32.90 | **25.81** | 0.031 | 0.033 | 0.035 |
| 3 | 38.40 | 37.41 | **30.39** | 0.041 | 0.041 | 0.043 |
| 4 | 36.50 | 35.65 | **27.73** | 0.038 | 0.038 | 0.041 |
| 5 | 32.06 | 32.10 | **24.07** | 0.029 | 0.030 | 0.033 |
| 6 | 18.14 | 32.65 | **26.68** | 0.024 | 0.025 | 0.031 |

**Fig. 10** shows the parity plot and relative error distribution of the fifth fold. The mean error is close to zero (0.01%), indicating that the model does not exhibit significant systematic bias. Additionally, the MAPE is only 3.30%, demonstrating high predictive accuracy. The error distribution exhibits a clear normal concentration trend, suggesting that most predicted values deviate only slightly from the true values, with only a few data points showing larger deviations.

This indicates that the model achieves good fit and low error in the target property prediction, exhibiting strong generalization capability.

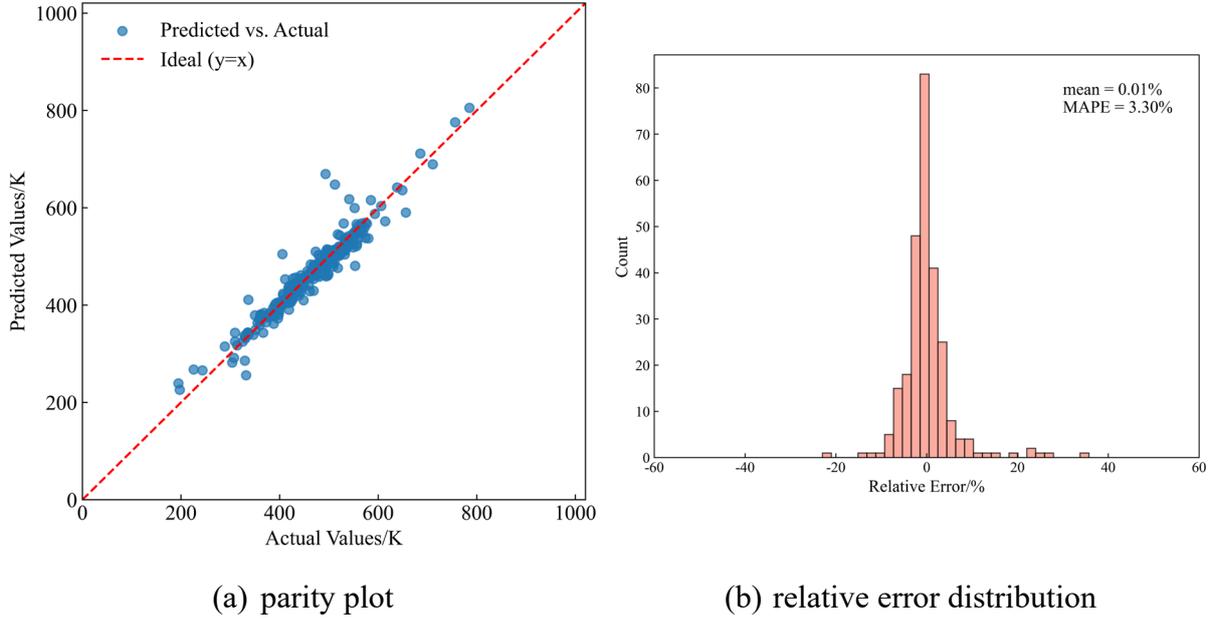

(a) parity plot  (b) relative error distribution

**Fig. 10.** Parity plot and relative error distribution of the fifth fold for $T_b$ prediction.

For $L_{mv}$, $T_c$ and $P_c$, we employ the same modeling approach as for $T_b$, using the six-fold cross validation. The results are presented in **Tables S1 – S3** of the **Supplementary Material**. We show the fold with the lowest RMSE on the test set among the six folds as the optimal one. The results are shown in **Table 3**, and the corresponding parity plots and error distribution plots are also produced, as shown in **Figs. S3–S5** of the **Supplementary Materials**.

**Table 3** Best fold results for $L_{mv}$, $T_c$, and $P_c$ using ANN (feature number = 160)

|  | RMSE | | | MAPE | | | $R^2$ |
| --- | --- | --- | --- | --- | --- | --- | --- |
|  | Training | Validation | Test | Training | Validation | Test | (testing set) |
| $L_{mv}$ | 0.0079 | 0.0098 | 0.010 | 0.048 | 0.057 | 0.052 | 0.99 |
| $T_c$ | 37.12 | 46.80 | 40.54 | 0.035 | 0.044 | 0.037 | 0.82 |
| $P_c$ | 2.72 | 5.32 | 3.32 | 0.055 | 0.091 | 0.068 | 0.92 |

To further evaluate the impact of model architecture on the final prediction results, we explore a deeper Deep Neural Network (DNN) architecture consisting of six hidden layers, with 512, 256, 128, 64, 32, and 16 neurons, respectively. Each layer employs the ReLU activation function. The model is comprehensively evaluated using 6-fold cross-validation. To ensure a fair comparison, the same feature set was used as in the ANN model. The results are summarized in **Table 4**.

**Table 4** Six-fold cross-validation results of $T_b$ using DNN (feature number =160)

| Fold | RMSE/K | | | MAPE | | |
| --- | --- | --- | --- | --- | --- | --- |
| | Training | Validation | Test | Training | Validation | Test |
| 1 | 36.54 | 44.11 | 37.34 | 0.06 | 0.06 | 0.06 |
| 2 | 20.92 | 33.63 | 28.90 | 0.03 | 0.05 | 0.04 |
| 3 | 22.52 | 27.05 | 28.24 | 0.02 | 0.04 | 0.04 |
| 4 | 19.84 | 24.45 | 28.32 | 0.03 | 0.03 | 0.04 |
| 5 | 20.41 | 28.07 | 28.97 | 0.03 | 0.04 | 0.04 |
| 6 | 20.23 | 68.83 | 32.78 | 0.03 | 0.04 | 0.04 |

The results indicate that, given the feature representation and dataset size used in this study, the DNN model does not significantly outperform ANN. As shown in **Table 4**, the best-performing fold is fold 3, with RMSE of 28.24 K on the test set, which is notably higher than the best-performing fold in the ANN model (i.e., fold 5) with an RMSE of 24.07 K, as shown in **Table 2**. Furthermore, we investigate the impact of the number of DNN layers on model performance. The corresponding results are presented in **Table 5**. The hidden layer configuration of 512, 256, 128, 64, 32, and 16 indicates a total of six hidden layers, with each number representing the number of neurons in each layer.

**Table 5** Impact of number of hidden layers on prediction performance (feature number =160)

| Number of hidden layers | Hidden layer configuration | RMSE on test set from the best fold/K |
| --- | --- | --- |
| 6 | (512,256,128,64,32,16) | 28.24 |
| 5 | (512,256,128,64,32) | 28.13 |
| 4 | (512,256,128,64) | 28.42 |
| 3 | (512,256,128) | 27.32 |
| 2 | (512,256) | 25.37 |
| 1 | (512) | 25.09 |

It can be observed that as the number of DNN layers increases, the model performance gradually deteriorates. This is because deeper networks introduce a significantly larger number of parameters, increasing the difficulty of optimization. To maintain generalization without overfitting, more training samples are required to effectively learn the underlying features. Therefore, we ultimately selected the more compact ANN model— which demonstrated higher prediction accuracy on this dataset— as the primary model for this study.

Furthermore, we analyze the effect of the number of neurons on the results. The number of neurons increased from 50 to 600, and six-fold cross-validation was performed for each setting. The optimal fold was selected, and the results are shown in **Table 6**. It can be seen that the best prediction performance is achieved when the number of neurons was 500. Therefore, we ultimately selected an ANN model with one hidden layer containing 500 neurons.

**Table 6** Effect of the number of neurons on prediction performance (feature number = 160)

| No. of neurons | RMSE on test set from the best fold/K |
| --- | --- |
| 50 | 37.12 |
| 100 | 31.71 |
| 200 | 30.82 |
| 300 | 28.49 |
| 400 | 25.49 |
| 500 | 24.07 |
| 600 | 26.54 |

To evaluate the superiority of the features generated in this work to those generated using GC by Alshehri et al. [10], we use the features generated by GC as inputs to an ANN model. For a fair comparison, we increase the number of features established by connectivity matrix in this study from 160 to 424, having the same number of features as Alshehri et al.[10]. The resulting ANN model is referred to as ANN–GC. Furthermore, a GPR model using features derived from the connectivity matrix is also trained, with its structure and hyperparameters identical to those of Alshehri et al [10], and with a molecular feature dimension of 424. The resulting GPR model is denoted as GPR–CM. Data splitting for these two models is the same as [10], namely training and validation: testing = 95%: 5%. The ANN model established using our CM features is referred to as ANN–CM, while the GPR model using GC features from Alshehri et al. [10] is designated as GPR–GC.

The comparative results from ANN–CM and ANN–GC are shown in **Table 7**, where the optimal RMSE is 36.04 and $R^2$ is 0.81. Under the same number of features, the proposed framework achieves an RMSE of 25.15 and $R^2$ of 0.91, indicating that the ANN model developed in this study effectively learns the mapping relationships between different features and target properties and thus demonstrating strong generalization capability.

Table 7 Comparative results from ANN-CM and ANN-GC on test set

| Fold | RMSE/K | | | $R^2$ | | |
|---|---|---|---|---|---|---|
| | ANN-CM | ANN-GC | Improved | ANN-CM | ANN-GC | Improved |
| 1 | **30.97** | 37.76 | 17.98% | **0.86** | 0.79 | 8.35% |
| 2 | **28.56** | 39.69 | 28.03% | **0.88** | 0.77 | 13.98% |
| 3 | **29.33** | 36.04 | 18.60% | **0.87** | 0.81 | 7.67% |
| 4 | **29.79** | 38.97 | 23.55% | **0.87** | 0.78 | 11.49% |
| 5 | **25.15** | 37.35 | 32.66% | **0.91** | 0.80 | 13.57% |
| 6 | **30.36** | 38.04 | 20.19% | **0.86** | 0.79 | 9.45% |

The performance of the GPR–GC model from Cao et al [17] and Alshehri et al. [61] for predicting $T_b$, as well as the GPR-WP-GC [17] and GPR–CM models, is given in **Table 8**. The performance of these models for predicting $L_{mv}$, $T_c$, and $P_c$ is presented in **Tables 9–11**., with a molecular feature dimension of 424. The results shown Tables 8-11 correspond to the fold with the lowest test RMSE in the 20-fold cross-validation, with all hyperparameters and data splitting ratios kept consistent. The complete results of 20 folders from the GPR–CM model for predicting the four properties are provided in **Figs. S4–S7** of the **Supplementary Materials**.

Table 8 Comparative results for $T_b$ prediction using GPR model with different features

| Model | All dataset | | Test dataset | |
|---|---|---|---|---|
| | $R^2$ | RMSE/K | $R^2$ | RMSE/K |
| GPR-GC-Cao [17] | 0.99 | 7.65 [17][η] | 0.86 | 28.29 [17][η] |
| GPR-GC- Alshehri [61] | 0.95 | 10.71 [61][η] | 0.92 | 22.00 |
| GPR-CM (this work) | **0.99** | 8.99 | **0.94** | **20.24** |
| GPR-WP-GC [17] | 0.99 | **6.88 [17][η]** | 0.89 | 25.39 [17][η] |

[η] results directly taken from reference.

Table 9 Comparative results for $L_{mv}$ prediction using GPR model in different features

| Model | All dataset | | Test dataset | |
|---|---|---|---|---|
| | $R^2$ | RMSE/cc·mol$^{-1}$ | $R^2$ | RMSE/cc·mol$^{-1}$ |
| GPR–GC-Cao [17] | 0.999 | 0.0020[17] | 0.98 | 0.009[17] |
| GPR–GC- Alshehri [61] | 0.999 | 0.00[61] | 0.99 | 0.007 |
| GPR–CM (this work) | 0.999 | **0.0008** | 0.99 | **0.003** |
| GPR–WP–GC [17] | 1.000 | 0.0020[17] | 0.99 | 0.007[17] |

**Table 10** Comparative results for $T_c$ prediction using GPR model in different features

| Model | All dataset | | Test dataset | |
|---|---|---|---|---|
| | $R^2$ | RMSE/ K | $R^2$ | RMSE/K |
| GPR–GC-Cao [17] | 0.94 | 24.30[17] | 0.37 | 91.14[17] |
| GPR–GC- Alshehri [61] | 0.99 | 11.04[61] | 0.97 | 17.96 |
| **GPR–CM (this work)** | **0.99** | **4.40** | **0.98** | **14.75** |
| GPR–WP–GC [17] | 0.93 | 23.15[17] | 0.43 | 86.80[17] |

**Table 11** Comparative results for $P_c$ prediction using GPR model in different features

| Model | All dataset | | Test dataset | |
|---|---|---|---|---|
| | $R^2$ | RMSE/bar | $R^2$ | RMSE/bar |
| GPR–GC-Cao [17] | 0.99 | 0.77[17] | 0.87 | 2.87[17] |
| GPR–GC- Alshehri [61] | 0.96 | 1.42[61] | 0.96 | 1.94 |
| **GPR–CM (this work)** | **0.99** | **0.58** | **0.99** | **1.06** |
| GPR–WP–GC [17] | 0.99 | 0.74[17] | 0.88 | 2.77[17] |

The results in **Table 8–11** show that for the predictions of $L_{mv}$, $P_c$, and $T_c$, the GPR–CM model achieved the best performance both on the test sets and across the entire datasets. For the $T_b$ prediction, it achieved the best performance on the test set and ranked second across all data. This slight discrepancy is due to the difference in the number of molecules used: the GPR-WP-GC model [17] was trained on 4658 molecules, while the GPR–CM model in this work used a total of 5276 molecular samples. Since Cao et al.[17] did not disclose which molecules were excluded, it is not possible to provide the prediction results of our GPR–CM model on the exact same set of 4658 molecules. As a result, our RMSE (i..e, 8.99) is a bit higher than their results of 6.88 on the entire dataset. However, compared to the GPR–GC–Alshehri model [61], which was also evaluated using 5276 molecules, our model achieved better overall performance, with an RMSE of 8.99 versus 10.71.

To better evaluate the model's performance, the percentage deviation between predicted and measured property values is calculated using **Eq. (21)**, and the proportion of molecules falling within percentage deviation thresholds of 1%, 5%, and 15% is determined. The results show that 37.91% of molecules have a relative absolute percentage error (RAPE) below 1%, while 90.45%, 97.23%, and 98.77% fall under 5%, 10%, and 15%, respectively, as illustrated in **Fig. 11**. The comparative results with the mainstream machine learning models [17] are presented in **Table 12**.

$$RAE(\%) = 100 \times |y_{pred} - y_{true}|/y_{true} \tag{21}$$

**Table 12** Comparative results with mainstream machine learning models for $T_b$ prediction.

| Model | 1% error | 5% error | 10% error | 15% error |
|---|---|---|---|---|
| SVR | 30.98% | 80.36% | 92.53% | - |
| LightGBM | 20.55% | 68.61% | 86.41% | - |
| BP | 34.16% | 86.88% | 96.31% | - |
| CNN | 18.38% | 71.75% | 91.41% | - |
| ANN – CM | 37.9% | 90.4% | 97.2% | 98.7 |
| GPR – CM | 92.57% | 99.00% | 99.77% | 99.87% |
| GPR – GC | 79.1% | 87.8% | 92.7% | -- |

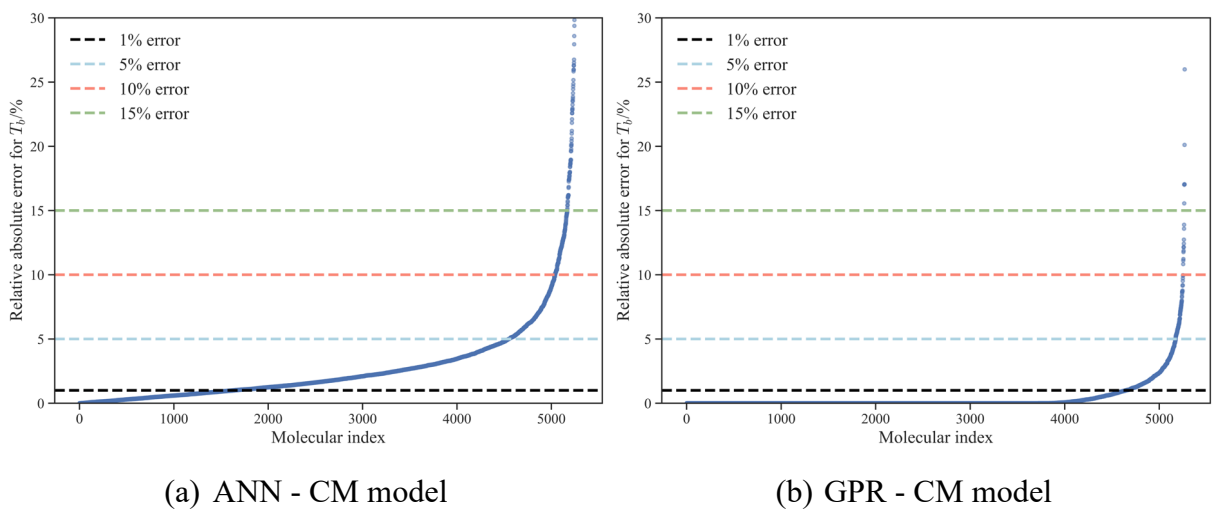

(a) ANN - CM model  (b) GPR - CM model

**Fig. 11.** RAE plots for $T_b$ prediction on the entire dataset.

The results in **Table 12** indicate that the proposed model outperforms the currently mainstream machine learning models at deviation thresholds of 1%, 5%, and 15%. From **Fig. 11 (a)**, it can be observed that the prediction error distribution for most molecules is relatively smooth, without sudden peaks of exceptionally high errors in the middle. This indicates that the model has learned a continuous and stable mapping relationship, allowing to maintain a relatively controlled error range when predicting previously unseen molecules from the training set, demonstrating strong generalization capability.

Conversely, if the model merely memorizes specific molecules from the training set or overfitting, the error would likely spike sharply when testing new or slightly different molecules, leading to abrupt jumps in prediction accuracy, like **Fig. 11(b)**, the prediction error remains extremely low for the first 4000 molecules but increases sharply once the number of molecules reaches approximately 4500. This observation leads us to preliminarily infer that the GPR-based prediction model may suffer from overfitting. In machine learning, overfitting typically manifests as significantly better performance on the training set compared to the test set [62]. This phenomenon is observed both in **Table 8** and in the results reported by Cao et al. [17]. Consequently, evaluating model performance solely based on the entire dataset may create a misleading impression of exceptionally low error, resulting in an overestimation of the model's generalization ability. Therefore, we shifted our evaluation focus to the model's performance on the test set rather than on the entire dataset. The results are presented in **Table 13**. It can be observed that for ANN-CM, the $R^2$ on the test set for all four properties are the highest. Although the $R^2$ for $P_c$ and $L_{mv}$ are also superior compared to other models, the RMSE are slightly higher. This slight discrepancy may be attributed to differences in sample sizes. Although the same database was used in this work as well as in GPR-GC and GPR-WP, there are deviations in the number of samples. In this work, the sample sizes for $T_b$, $P_c$, $T_c$, and $L_{mv}$ are 5276, 775, 777, and 1057, respectively, whereas in their studies, the corresponding numbers are 4658, 717, 724, and 987 [17]. Since the method for removing missing molecules was not disclosed in their work, we adopted a more comprehensive original dataset in this study.

Table 13 The performance of different models on the test set

| Properties | Performance indicator | ANN-CM | GRP-GC | GPR-WP |
|---|---|---|---|---|
| $T_b$ | $R^2$ | **0.91** | 0.86 | 0.89 |
|  | RMSE/K | **24.07** | 28.29 | 25.39 |
| $P_c$ | $R^2$ | **0.92** | 0.87 | 0.88 |
|  | RMSE/bar | 3.32 | 2.87 | **2.77** |
| $T_c$ | $R^2$ | **0.82** | 0.37 | 0.43 |
|  | RMSE/K | **40.54** | 91.14 | 86.80 |
| $L_{mv}$ | $R^2$ | **0.99** | 0.98 | 0.99 |
|  | RMSE/cc·mol$^{-1}$ | 0.010 | 0.009 | **0.007** |

In summary, the connectivity matrix-based molecular representation method proposed in this work demonstrates superior predictive performance compared to the mainstream GC-based machine learning methods when both approaches utilize a 424-dimensional feature set.

### 5.3 Model interpretability analysis

To further enhance the interpretability of the model, Shapley values are employed to quantify the contribution of each feature to the model's prediction, as shown in **Fig. 12**. This figure illustrates the relationship between molecular features and their corresponding Shapley values. The points in the plot are color-coded, where red represents higher feature values and blue represents lower feature values. The Shapley value of zero serves as the reference line, with points on the left indicating a negative contribution to the predicted target value and points on the right indicating a positive contribution. Taking feature 1st_153 as an example, its higher feature values are predominantly located in the right half of the plot, indicating that the more frequently this feature appears in a molecule, the higher the predicted boiling point.

For $T_b$ prediction, features such as 1st_153, 1st_5 and 1st_132 exhibit relatively higher contributions to the model's prediction, and show a positive correlation. Notably, 1st_153 consistently demonstrates the most significant influence across nearly all predictions.

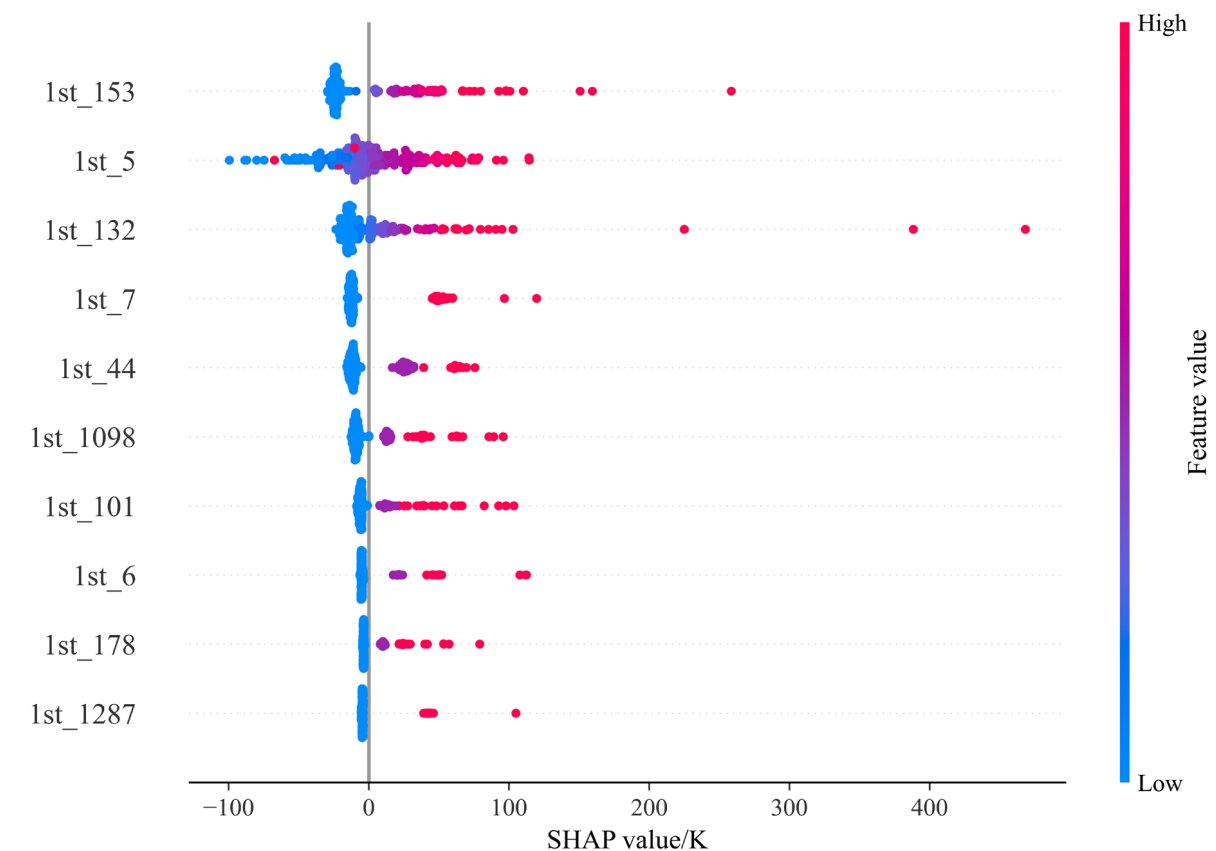

**Fig. 12.** Shapley values of the top 20 features for $T_b$ prediction

Referring to our feature definition rules, 1st_153 represents the first-order feature (determinant = 153) corresponding to the first-order environment of carbon atoms on a benzene ring, as illustrated in **Fig. 13**. This means that the greater the number of benzene rings, the higher the $T_b$ value, which is consistent with the mechanistic explanation. The number of benzene rings is proportional to the molecular mass, and as the molecular mass increases, more energy is required to overcome intermolecular forces, leading to a higher boiling point [63].

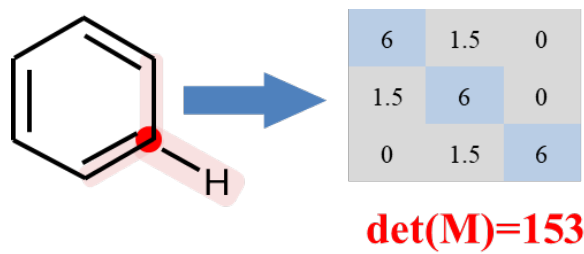

**Fig. 13.** The molecular structure corresponding to 1st_153

Similarly, 1st_132 represents the first-order feature (determinant = 132) of the first-order environment of the methylene carbon in propane as shown in **Fig. 14,** in this context, R does not represent a specific atom or group; rather, it is a placeholder used to simplify the structure and indicate a general part of the molecule that does not affect the core functional group or skeleton under discussion. This means that the more frequently this feature appears, the longer the carbon chain, leading to a higher boiling point, which aligns with the mechanistic theory, similarly, the same applies to other highly contributing features. We further established the correspondence between molecular structures and features, as shown in **Table 15**, for reference.

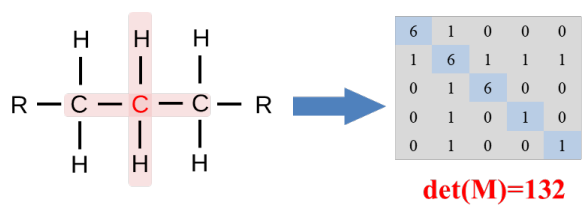

**Fig. 14.** The molecular structure corresponding to 1st_132

**Table 15** Correspondence between molecular structures and features

| Feature name | Atomic environment | |
|---|---|---|
| 1st_5 | The hydrogen atom is bonded to a carbon atom | C — H |
| 1st_44 | The carbon atom and the oxygen atom are bonded by a double bond. | C=O |
| 1st_153 | The first-order environment of any carbon atom on the benzene ring | |
| 1st_1307 | The first-order environment of the methine carbon in cyclohexane | |
| 1st_17 | First-order environment of the methylene carbon | |
| 1st_132 | The first-order environment of the methylene carbon in propane | 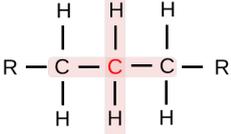 |
| 1st_41 | The first-order environment of the oxygen atom in a hydroxyl group | 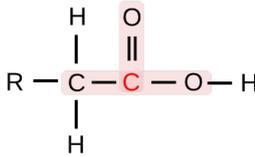 |
| 1st_101 | The bonding structure between chlorine and carbon atoms | 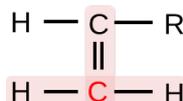 |
| 1st_2000 | The first-order environment of the carbon atom in a carboxylic acid functional group | 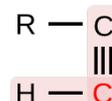 |
| 1st_20 | The first-order environment of the methylene carbon in a vinyl group | 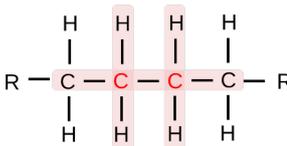 |
| 1st_21 | The first-order environment of the alkyne carbon in an alkynyl group | 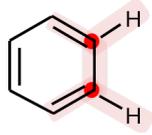 |
| 2nd_493 | The atomic environment of the two connected methylene carbons in n-butane | 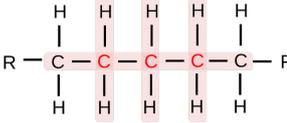 |
| 2nd_689 | The atomic environment of two adjacent carbon atoms on the benzene ring | 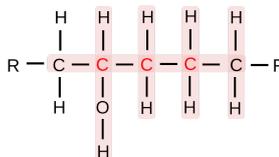 |
| 3rd_1840 | The atomic environment of the four methylene carbons in n- pentane backbone | |
| 3rd_15976 | The atomic environment of the methine carbon bonded to two methylene carbons in 2-pentanol | |
| 4th_6867 | The atomic environment of the four methylene carbons in n-hexyl backbone | 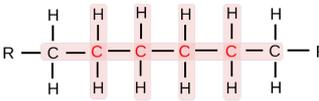 |

To investigate the influence of molecular structure on different properties, the Shapley values for $P_c$ are also calculated. The results are depicted in **Fig. 15**. It can be observed that for features 2nd_493 and 4th_6867, higher occurrence frequencies correspond to lower Shapley values, indicating a negative correlation. The mechanistic meaning of 2nd_493 is the atomic environment of the two connected methylene carbons in n-butane, while 4th_6867 represents the atomic environment of the four methylene carbons in the n-hexyl backbone. A higher frequency of these features implies a longer carbon chain, which corresponds to a lower value of $P_c$. From a mechanistic perspective, as the carbon chain lengthens, the van der Waals forces between molecules increase, making it easier for the molecules to condense into a liquid at lower pressures—an effect that can also be reasonably explained by the van der Waals equation.

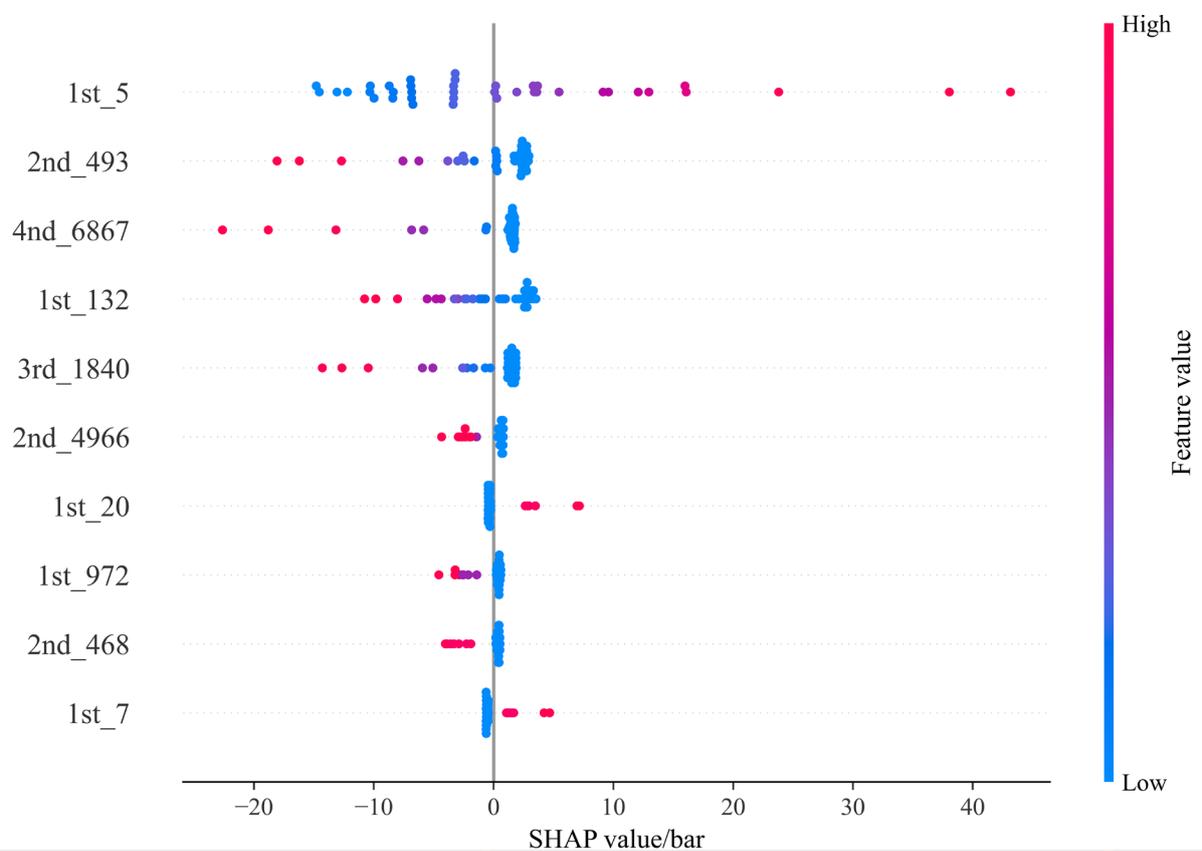

**Fig. 15.** Shapley values of the top 20 features for $P_c$ prediction

We also calculated the normalized Shapley values for $T_c$ and plotted them together with the results for $T_b$ in a figure, a noticeable overlap in the feature importance is observed between the predictions of $T_b$ and $T_c$, as shown in **Fig. 16,** the Shapley values for $L_{mv}$ can be found in **Fig. S6** of the **Supplemental Materials.** Features such as 1st_5, 1st_132, 1st_153, 4th_6867, 1st_41, and 2nd_689 exhibit similar contribution rankings in both models, indicating that these structural characteristics influence both properties in comparable ways. The high-frequency features mentioned above can be categorized into those that influence the $T_c$ and $T_b$ of linear alkanes, such as 1st_5, 1st_132, and 4th_6867. Their frequent occurrence indicates an increase in carbon chain length. As analysed above, the increase in carbon chain length not only determines the magnitude of $T_b$ but also affects $T_c$. With longer carbon chains, more energy is required for molecules to overcome intermolecular attractions and transition into the gas phase, resulting in a higher $T_c$.

Based on the above analysis, we can conclude that $T_c$ and $T_b$ exhibit a positive correlation.

Furthermore, we plotted the predicted values of $T_b$ and $T_c$ in ascending order, as shown in **Fig. S7** of the **Supplemental Materials**. The results show that across the entire dataset, $T_c$ and $P_c$ exhibit a proportional relationship, and $T_c$ is consistently higher than $T_b$. To enable a more intuitive comparison, we further plotted a parity plot as shown in **Fig. S8** of the **Supplemental Materials**. It can be seen that all samples in the upper-left region, except for two points that appear slightly above the line. These two samples are C1CCCCCC1 (cyclooctane) and C1CCCCCCC1 (cyclononane), with predicted $T_b$ values of 408.31 K and 430.12 K, and corresponding predicted $T_c$ values of 416.91 K and 433.38 K, respectively.

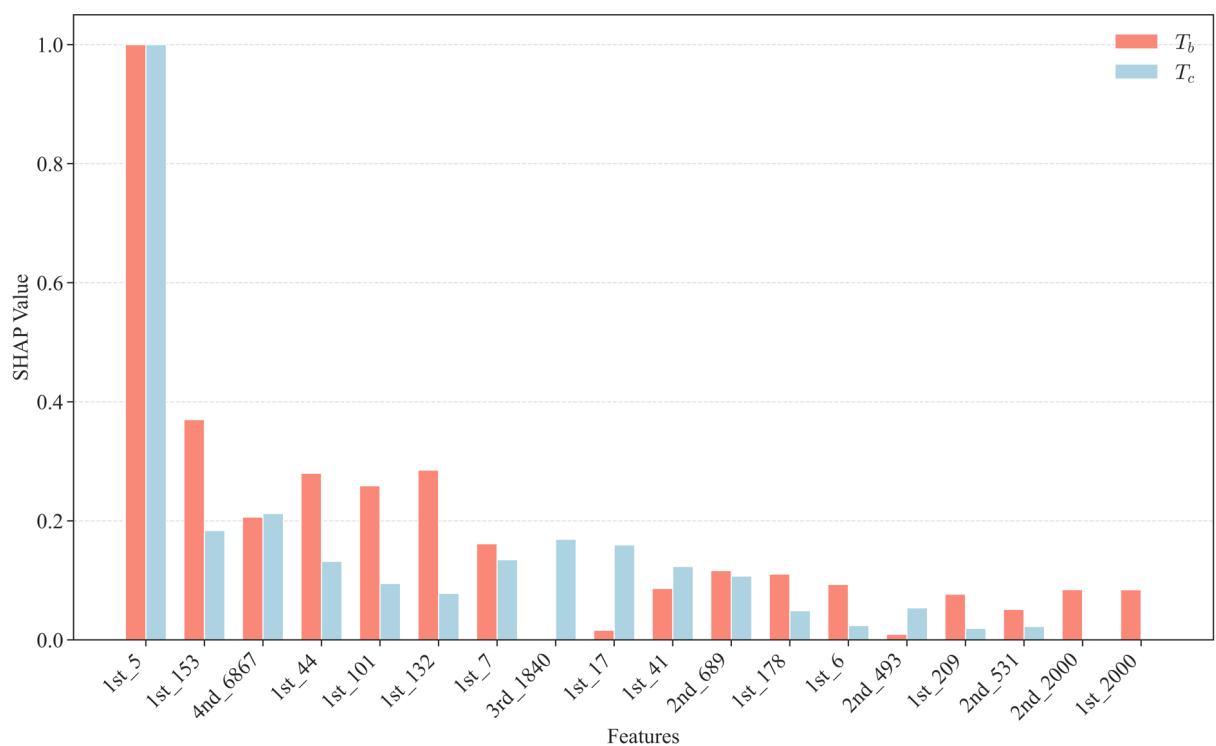

**Fig. 16.** Shapley values for $T_b$ and $T_c$

We also calculated the normalized Shapley values for $P_c$ and plotted them together with the results for $T_b$, as shown in **Fig. 17**. In contrast, certain features such as 2nd_493 and 1st_972, rank much lower in the prediction of $T_b$ and $T_c$, suggesting that $P_c$ responds differently to molecular structural variations. Notably, 1st_5 consistently demonstrates the most significant influence across nearly all predictions, highlighting its critical role in property estimation. This feature represents the local structural environment where a hydrogen atom is directly bonded to a carbon atom, shown in **Table 12**.

From a mechanistic perspective, a greater number of C–H bonds increases the molecule's surface area, which in turn leads to a higher boiling point due to stronger intermolecular forces. These enhanced interactions make it more difficult for the molecule to transition into the gas phase, thereby resulting in an elevated $T_c$. Additionally, the spatial arrangement of C–H bonds affects the molecule's volume and packing efficiency. Molecules with smaller volumes and tighter packing—such as ortho-xylene—require higher external pressure to condense into the liquid phase during gas-liquid phase transitions, resulting in a higher $P_c$. Therefore, the local structural feature of hydrogen atoms directly attached to carbon atoms has a direct and significant impact on the prediction accuracy of $T_b$, $P_c$, $T_c$.

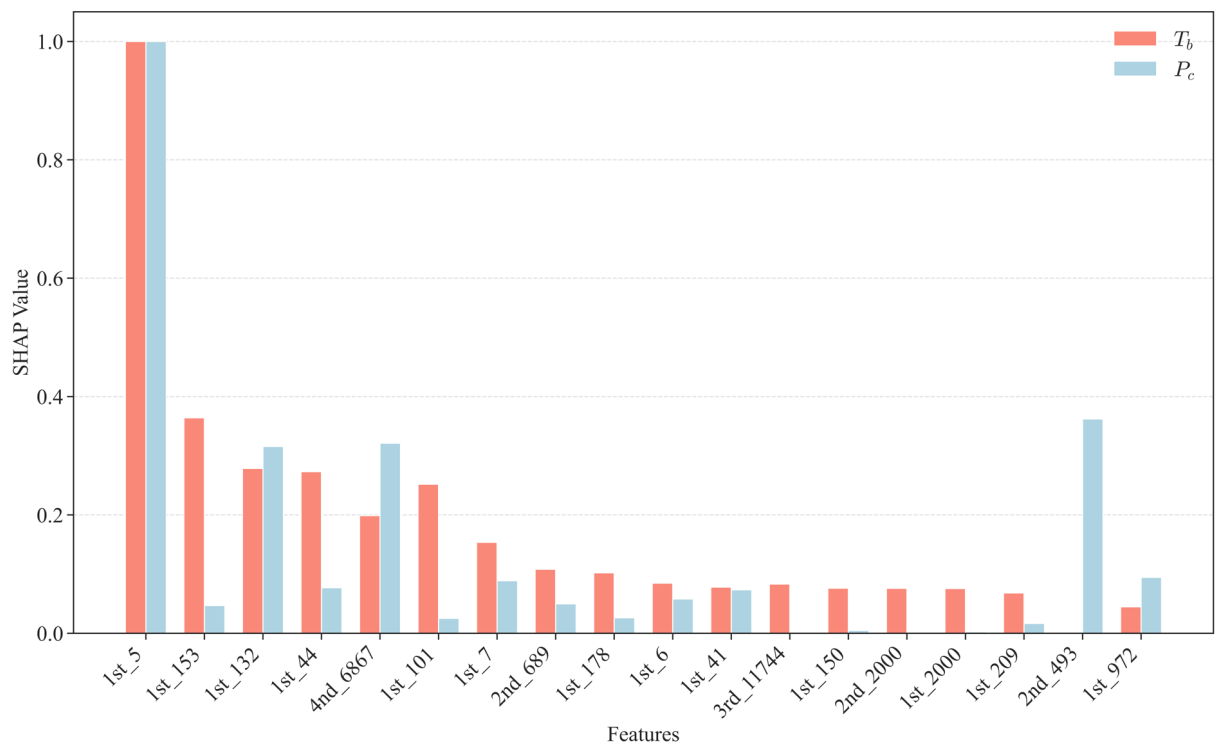

**Fig. 17.** Shapley values for $T_b$ and $P_c$.

## 6 Conclusion

Accurate prediction of molecular properties is crucial for product design and chemical process modelling and optimization. This study proposed a pure component property estimation framework using explainable machine learning methods. The molecular representation method based on the connectivity matrix effectively considered atomic bonding relationships to automatically generate features. Moreover, the one-to-one mapping between molecular features and labels was improved, with the joint conditional entropy reduced by 29% compared to the dataset built using the GC method. Furthermore, feature importance was computed using the RF model for feature pooling while preserving the original physical meaning of the features. The ranked important features served as inputs for the property prediction model. In this study, $T_b$ $L_{mv}$, $T_c$ and $P_c$ were predicted, the results demonstrated that the proposed method outperformed the current mainstream machine learning models. The prediction results indicate that when the descriptor count is fixed at 424—the same as in the GC approach—the GPR-CM model outperforms the GC-based GPR-GC and GPR-WP models in every prediction task. On the test set, its RMSE drops by as much as 39.7 % for $T_b$ prediction, 66.7 % for $L_{mv}$ prediction, 83.8 % for $T_c$ prediction, and 63.0 % for $P_c$ prediction. Subsequent analysis showed that GPR tends to over-fit, so a more extrapolative ANN-CM model was adopted. For the four target properties, ANN-CM matches GPR-WP on $L_{mv}$ while boosting the test-set $R^2$ by 1 % compared with GPR-GC; for the other three properties it surpasses both GPR-WP and GPR-GC on the test set, thus, for all four property prediction tasks, the results obtained with both GPR models and ANN models demonstrate that the CM based molecular feature extraction method proposed in this study outperforms the GC approach. Finally, Shapley values were employed for model interpretability analysis, identifying key molecular structures affecting the property, such as 1st_153, 1st_5, and 1st_132. The correspondence between common molecular structures and features is also provided. For example, 1st_132 represents the first-order environment of the methylene carbon in propane, and 4th_6867 represents the atomic environment of the four methylene carbons in the n-hexyl backbone. The study also confirmed that different molecular properties were influenced by different structural features, aligning with mechanistic interpretations. In conclusion, the proposed framework was demonstrated to be feasible and provided a solid foundation for molecular-level management in chemical processes.

In the future, thanks to the interpretability of the proposed framework, the key molecular structures influencing different properties can be systematically identified, facilitating the discovery of common patterns among them. A novel property prediction approach based on transfer learning will be developed to enable large-scale prediction of additional pure component properties. Moreover, the proposed framework will be further applied to tasks such as mixture component reconstruction and process integration modelling to achieve multi-scale process optimization.

**Acknowledgement**

The authors gratefully acknowledge the financial support from China Scholarship Council (CSC) (No. 202406440073).

**Declaration of Competing Interest**

The authors declare no competing financial interests.

**Authorship contribution**

Jianfeng Jiao: Writing – Original draft, Data curation, Formal analysis, Investigation, Methodology, Project administration, Software, Validation, Visualization; Xi Gao: Validation, writing – review & editing, Jie Li: Writing – review & editing, Conceptualization, Methodology, Funding acquisition, Resources, Validation, Supervision.

**Abbreviations**

| | List of Acronyms and Abbreviations |
|---|---|
| ANN | Artificial neural networks |
| ARE | Absolute relative error |
| Bi-GRU | Bidirectional gated recurrent unit |
| CAMD | Computer-aided molecular design |
| CNN | Convolutional neural network |
| CM | Connectivity matrices |
| ECFPs | Extended-Connectivity Fingerprints |
| FCC | Fluid catalytic cracking |
| GC | Group contribution |
| GMM | Gaussian Mixture Models |
| GPR | Gaussian process regression |
| GNN | Graph neural networks |
| KDE | Kernel density estimation |
| MAE | Mean Absolute Error |
| MAPE | Mean absolute percentage error |
| NMR | Nuclear magnetic resonance |
| PCA | Principal component analysis |
| PDF | Probability density functions |
| RF | Random forest |
| RNNs | Recurrent neural network |
| RMSE | Root mean square error |
| RAPE | Relative absolute percentage error |
| SMILES | Simplified molecular-input line-entry system |
| SVR | Support Vector Regression |

**References**

[1] P.J. Morris, Wetter is better for peat carbon, *Nat. Clim. Chang.* 11 (2021) 561–562.

[2] Y. Yang, B. Chen, PSE in China: retrospect and prospects, *Huagong Jinzhan/Chemical Ind. Eng. Prog.* 41 (2022) 3991–4008.

[3] Wu Y, Molecular management for refining operations, University of Manchester, 2010.

[4] D. Guan, L. Zhang, Initial guess estimation and fast solving of petroleum complex molecular reconstruction model, *AIChE J.* 68 (2022) e17782.

[5] H. Gu, J. Li, P. Mu, Q. Zhu, Improving the Operational Efficiency of Ethylene Cracking Integrated with Refining by Molecular Management, *Ind. Eng. Chem. Res.* 59 (2020) 13160–13174.

[6] E.T.C. Vogt, B.M. Weckhuysen, The refinery of the future, *Nature* 629 (2024) 295–306.

[7] V. Mann, R. Gani, V. Venkatasubramanian, Group contribution-based property modeling for chemical product design: A perspective in the AI era, *Fluid Phase Equilib.* 568 (2023) 113734.

[8] Y. Wei, L. Shan, T. Qiu, D. Lu, Z. Liu, Machine learning-assisted retrosynthesis planning: Current status and future prospects, *Chinese J. Chem. Eng.* 77 (2025) 273–292.

[9] J. Marrero, R. Gani, Group-contribution based estimation of pure component properties, *Fluid Phase Equilib.* 183–184 (2001) 183–208.

[10] A.S. Alshehri, A.K. Tula, F. You, R. Gani, Next generation pure component property estimation models: With and without machine learning techniques, *AIChE J.* 68 (2022).

[11] R. Gani, Group contribution-based property estimation methods: advances and perspectives, *Curr. Opin. Chem. Eng.* 23 (2019) 184–196.

[12] A. Fredenslund, Vapor-liquid equilibria using UNIFAC: a group-contribution method, Elsevier, 2012.

[13] D.S. Abrams, J.M. Prausnitz, Statistical thermodynamics of liquid mixtures: A new expression for the excess Gibbs energy of partly or completely miscible systems, *AIChE J.* 21 (1975) 116–128.

[14] A.R.N. Aouichaoui, F. Fan, J. Abildskov, G. Sin, Application of interpretable group-embedded graph neural networks for pure compound properties, *Comput. Chem. Eng.* 176 (2023) 108291.

[15] J. Burger, V. Papaioannou, S. Gopinath, G. Jackson, A. Galindo, C.S. Adjiman, A hierarchical method to integrated solvent and process design of physical $CO_2$ absorption using the SAFT-γ Mie approach, *AIChE J.* 61 (2015) 3249–3269.

[16] T. Zhou, Z. Song, X. Zhang, R. Gani, K. Sundmacher, Optimal Solvent Design for Extractive Distillation Processes: A Multiobjective Optimization-Based Hierarchical Framework, *Ind. Eng. Chem. Res.* 58 (2019) 5777–5786.

[17] X. Cao, M. Gong, A. Tula, X. Chen, R. Gani, V. Venkatasubramanian, An Improved Machine Learning Model for Pure Component Property Estimation, *Engineering* 39 (2024) 61–73.

[18] A.R.N. Aouichaoui, F. Fan, S.S. Mansouri, J. Abildskov, G. Sin, Combining Group-Contribution Concept and Graph Neural Networks Toward Interpretable Molecular Property Models, *J. Chem. Inf. Model.* 63 (2023) 725–744.

[19] R. Gani, P.M. Harper, M. Hostrup, Automatic Creation of Missing Groups through Connectivity Index for Pure-Component Property Prediction, *Ind. Eng. Chem. Res.* 44 (2005) 7262–7269.

[20] D.P. Visco, R.S. Pophale, M.D. Rintoul, J.L. Faulon, Developing a methodology for an inverse quantitative structure-activity relationship using the signature molecular descriptor, *J. Mol. Graph. Model.* 20 (2002) 429–438.

[21] D. Rogers, M. Hahn, Extended-Connectivity Fingerprints, *J. Chem. Inf. Model.* 50 (2010) 742–754.

[22] M.R. Dobbelaere, Y. Ureel, F.H. Vermeire, L. Tomme, C. V. Stevens, K.M. Van Geem, Machine Learning for Physicochemical Property Prediction of Complex Hydrocarbon Mixtures, *Ind. Eng. Chem. Res.* 61 (2022) 8581–8594.

[23] W. Mi, H. Chen, D. Zhu, T. Zhang, F. Qian, Melting point prediction of organic molecules by deciphering the chemical structure into a natural language, *Chem. Commun.* 57 (2021) 2633–2636.

[24] Z. Zeng, Y. Yao, Z. Liu, M. Sun, A deep-learning system bridging molecule structure and biomedical text with comprehension comparable to human professionals, *Nat. Commun.* 13 (2022) 862.

[25] G.B. Goh, N.O. Hodas, C. Siegel, A. Vishnu, SMILES2Vec: An Interpretable General-Purpose Deep Neural Network for Predicting Chemical Properties, (2017).

[26] V. Mann, K. Brito, R. Gani, V. Venkatasubramanian, Hybrid, Interpretable Machine Learning for Thermodynamic Property Estimation using Grammar2vec for Molecular Representation, *Fluid Phase Equilib.* 561 (2022) 113531.

[27] O. Wieder, S. Kohlbacher, M. Kuenemann, A. Garon, P. Ducrot, T. Seidel, T. Langer, A compact review of molecular property prediction with graph neural networks, *Drug Discov. Today Technol.* 37 (2020) 1–12.

[28]  B. Dou, Z. Zhu, E. Merkurjev, L. Ke, L. Chen, J. Jiang, Y. Zhu, J. Liu, B. Zhang, G.-W. Wei, Machine Learning Methods for Small Data Challenges in Molecular Science, *Chem. Rev.* 123 (2023) 8736–8780.

[29]  S. Ishida, T. Miyazaki, Y. Sugaya, S. Omachi, Graph Neural Networks with Multiple Feature Extraction Paths for Chemical Property Estimation, *Molecules* 26 (2021).

[30]  X. Zang, X. Zhao, B. Tang, Hierarchical Molecular Graph Self-Supervised Learning for property prediction, *Commun. Chem.* 6 (2023) 1–10.

[31]  S. Jain, B.C. Wallace, Attention is not explanation, *ArXiv Prepr. ArXiv1902.10186* (2019).

[32]  S. Wiegreffe, Y. Pinter, Attention is not not explanation, *ArXiv Prepr. ArXiv1908.04626* (2019).

[33]  B. Rozemberczki, L. Watson, P. Bayer, H.-T. Yang, O. Kiss, S. Nilsson, R. Sarkar, The shapley value in machine learning, in: 31st Int. Jt. Conf. Artif. Intell. 25th Eur. Conf. Artif. Intell., International Joint Conferences on Artificial Intelligence Organization, 2022: pp. 5572–5579.

[34]  A. Yang, S. Sun, Y. Su, Z.Y. Kong, J. Ren, W. Shen, Insight to the prediction of $CO_2$ solubility in ionic liquids based on the interpretable machine learning model, *Chem. Eng. Sci.* 297 (2024) 120266.

[35]  Q. Pan, X. Fan, J. Li, Automatic creation of molecular substructures for accurate estimation of pure component properties using connectivity matrices, *Chem. Eng. Sci.* 265 (2023) 118214.

[36]  K. Bi, T. Qiu, An intelligent SVM modeling process for crude oil properties prediction based on a hybrid GA-PSO method, *Chinese J. Chem. Eng.* 27 (2019) 1888–1894.

[37]  G.S.K. Ranjan, A.K. Verma, S. Radhika, K-Nearest Neighbors and Grid Search CV Based Real Time Fault Monitoring System for Industries, in: 2019 IEEE 5th Int. Conf. Converg. Technol., 2019: pp. 1–5.

[38]  Y. Wang, P. Wang, K. Tansey, J. Liu, B. Delaney, W. Quan, An interpretable approach combining Shapley additive explanations and LightGBM based on data augmentation for improving wheat yield estimates, *Comput. Electron. Agric.* 229 (2025) 109758.

[39]  Y. Shi, W. Zhong, X. Peng, M. Yang, W. Du, Interpretable reconstruction of naphtha components using property-based extreme gradient boosting and compositional-weighted Shapley additive explanation values, *Chem. Eng. Sci.* 284 (2024).

[40]  G. Landrum, Rdkit documentation, *Release* 1 (2013) 4.

[41]  M. Greenacre, P.J.F. Groenen, T. Hastie, A.I. d'Enza, A. Markos, E. Tuzhilina, Principal component analysis, *Nat. Rev. Methods Prim.* 2 (2022) 100.

[42]  Y. Mu, X. Liu, L. Wang, A Pearson's correlation coefficient based decision tree and its parallel implementation, *Inf. Sci. (Ny).* 435 (2018) 40–58.

[43]  S.J. Rigatti, Random Forest, *J. Insur. Med.* 47 (2017) 31–39.

[44]  M. Schonlau, R.Y. Zou, The random forest algorithm for statistical learning, *Stata J.* 20 (2020) 3–29.

[45]  F. Jin, Y. Guo, Y. Zhang, X. Ma, B. Liu, J. Jiao, X. Yin, H. Xu, J. Gong, K. Wen, A new gas pipeline network simulation method based on BHC-PINN, *Nat. Gas Ind.* 45 (2025) 164–174.

[46]  Z. Wang, D. He, H. Nie, Operational optimization of copper flotation process based on the weighted Gaussian process regression and index-oriented adaptive differential evolution algorithm, *Chinese J. Chem. Eng.* 66 (2024) 167–179.

[47]  A. Gulli, S. Pal, Deep learning with Keras, Packt Publishing Ltd, 2017.

[48]  B. Pang, E. Nijkamp, Y.N. Wu, Deep learning with tensorflow: A review, *J. Educ. Behav. Stat.* 45 (2020) 227–248.

[49]  X. Yin, K. Wen, Y. Wu, X. Han, Y. Mukhtar, J. Gong, A machine learning-based surrogate model for the rapid control of piping flow: Application to a natural gas flowmeter calibration system, *J. Nat. Gas Sci. Eng.* 98 (2022) 104384.

[50]  X. Yin, K. Wen, W. Huang, Y. Luo, Y. Ding, J. Gong, J. Gao, B. Hong, A high-accuracy online transient simulation framework of natural gas pipeline network by integrating physics-based and data-driven methods, *Appl. Energy* 333 (2023) 120615.

[51]  K. Wen, J. Jiao, K. Zhao, X. Yin, Y. Liu, J. Gong, C. Li, B. Hong, Rapid transient operation control method of natural gas pipeline networks based on user demand prediction, *Energy* 264 (2023) 126093.

[52]  O. Kramer, O. Kramer, Scikit-learn, *Mach. Learn. Evol. Strateg.* (2016) 45–53.

[53]  M. Sundararajan, A. Najmi, The many Shapley values for model explanation, in: Int. Conf. Mach. Learn., PMLR, 2020: pp. 9269–9278.

[54]  H. Cai, Y. Yang, Y. Tang, Z. Sun, W. Zhang, Shapley value-based class activation mapping for improved explainability in neural networks, *Vis. Comput.* (2025) 1–19.

[55]  G. Modla, Energy saving methods for the separation of a minimum boiling point azeotrope using an intermediate entrainer, *Energy* 50 (2013) 103–109.

[56] D.B. Robinson, D.Y. Peng, S.Y.K. Chung, The development of the Peng - Robinson equation and its application to phase equilibrium in a system containing methanol, *Fluid Phase Equilib.* 24 (1985) 25–41.

[57] L. Constantinou, R. Gani, J.P. O'Connell, Estimation of the acentric factor and the liquid molar volume at 298 K using a new group contribution method, *Fluid Phase Equilib.* 103 (1995) 11–22.

[58] Y.-C. Chen, A tutorial on kernel density estimation and recent advances, *Biostat. Epidemiol.* 1 (2017) 161–187.

[59] C.E. Shannon, A mathematical theory of communication, *Bell Syst. Tech. J.* 27 (1948) 379–423.

[60] L. Prechelt, Automatic early stopping using cross validation: quantifying the criteria, *Neural Networks* 11 (1998) 761–767.

[61] A.S. Alshehri, A.K. Tula, F. You, R. Gani, Corrections to "Next generation pure component property estimation models: With and without machine learning techniques," *AIChE J.* 69 (2023) e18086.

[62] Z.-H. Zhou, Machine learning, Springer nature, 2021.

[63] W.C. Edmister, Thermodynamic Properties of Hydrocarbons, *Ind. Eng. Chem.* 30 (1938) 352–358.

# Supplemental Materials for
# Pure Component Property Estimation Framework Using Explainable Machine Learning Methods

Jianfeng Jiao[1], Xi Gao[2,3,§] and Jie Li[1,*]

[1]Centre for Process Integration, Department of Chemical Engineering, School of Engineering, The University of Manchester, Manchester M13 9PL, UK

[2]School of Electronic and Information Engineering, Tongji University, Shanghai, China 201804

[3]School of Mechanical and Electrical Engineering, Jinggangshan University, Ji'an, Jiangxi, China 343009

**List of Tables**

**List of Figures**

[*] The corresponding author: Jie Li (jie.li-2@manchester.ac.uk). Tel: +44 (0) 161 529 3084

[§] The corresponding author: Xi Gao (gaoxi1979@126.com).

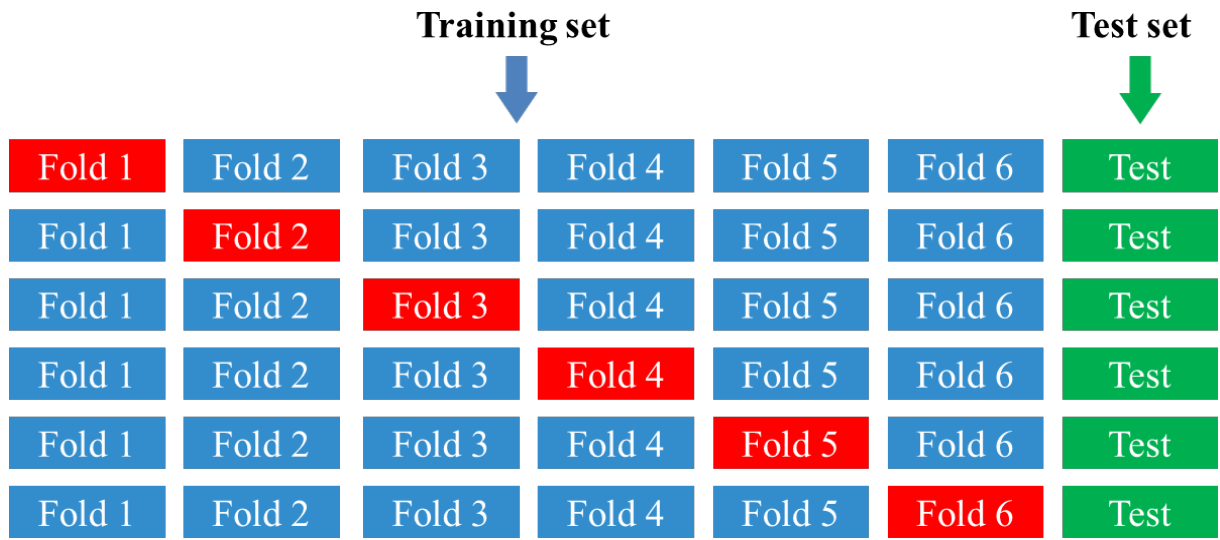

**Fig. S1.** Partition of training, validation and test sets.

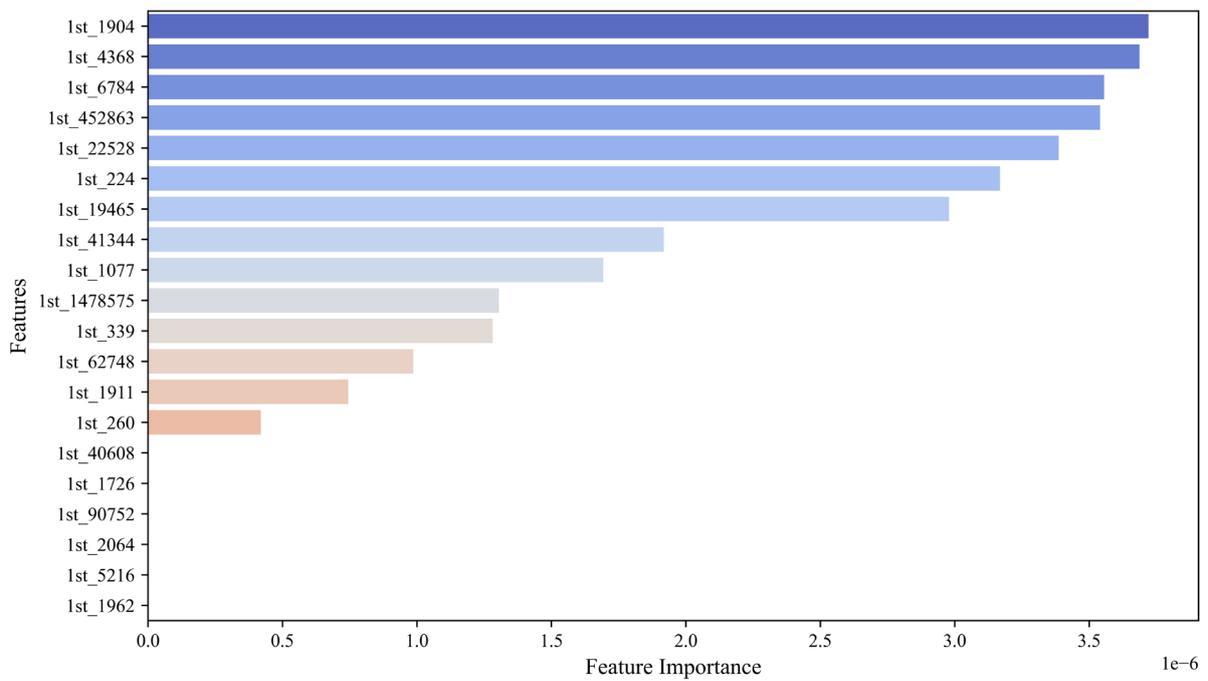

**Fig. S2.** Feature importance results of the first-order 240th-260th features for $T_b$ prediction

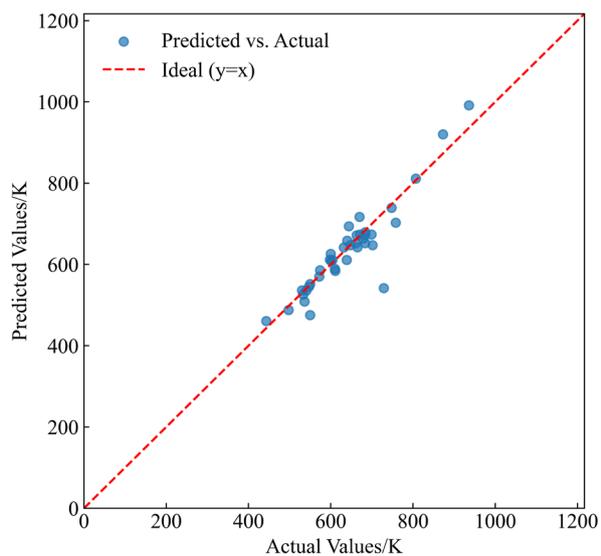 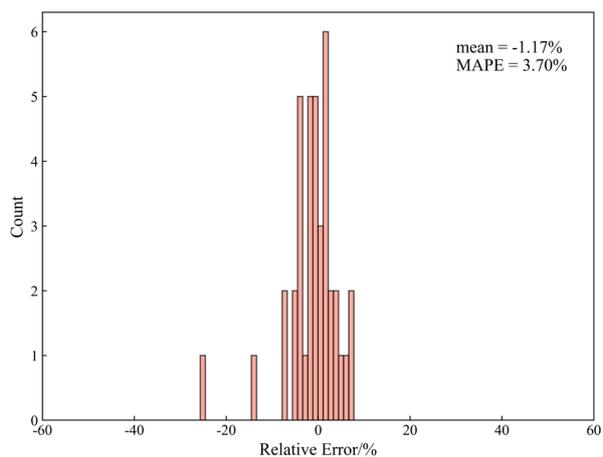

(a) parity plot

(b) relative error distribution

**Fig. S3.** Parity plot and relative error distribution for $T_c$ prediction.

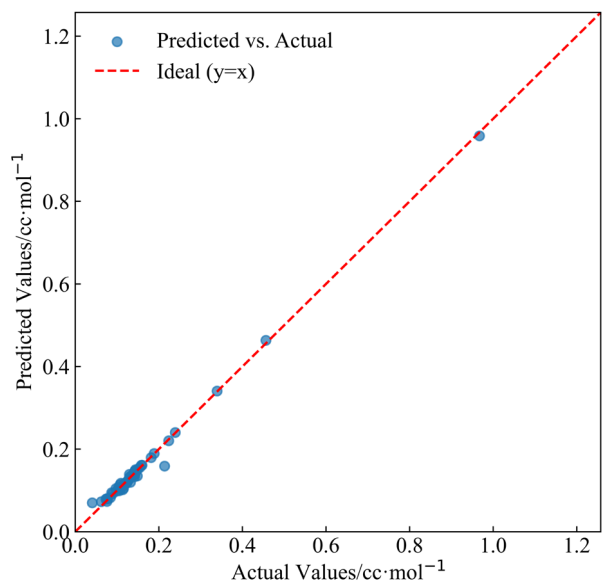 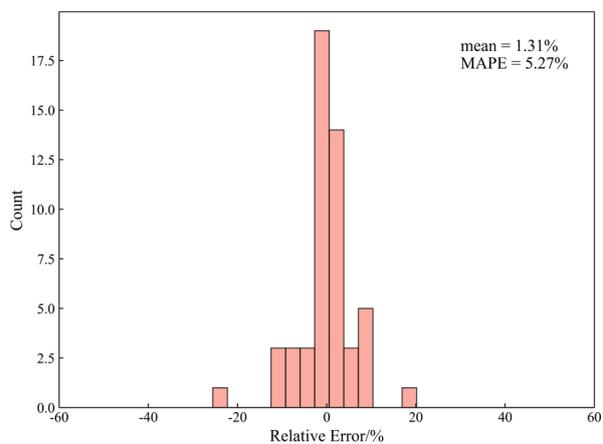

(a) parity plot

(b) relative error distribution

**Fig. S4.** Parity plot and relative error distribution for $L_{mv}$ prediction.

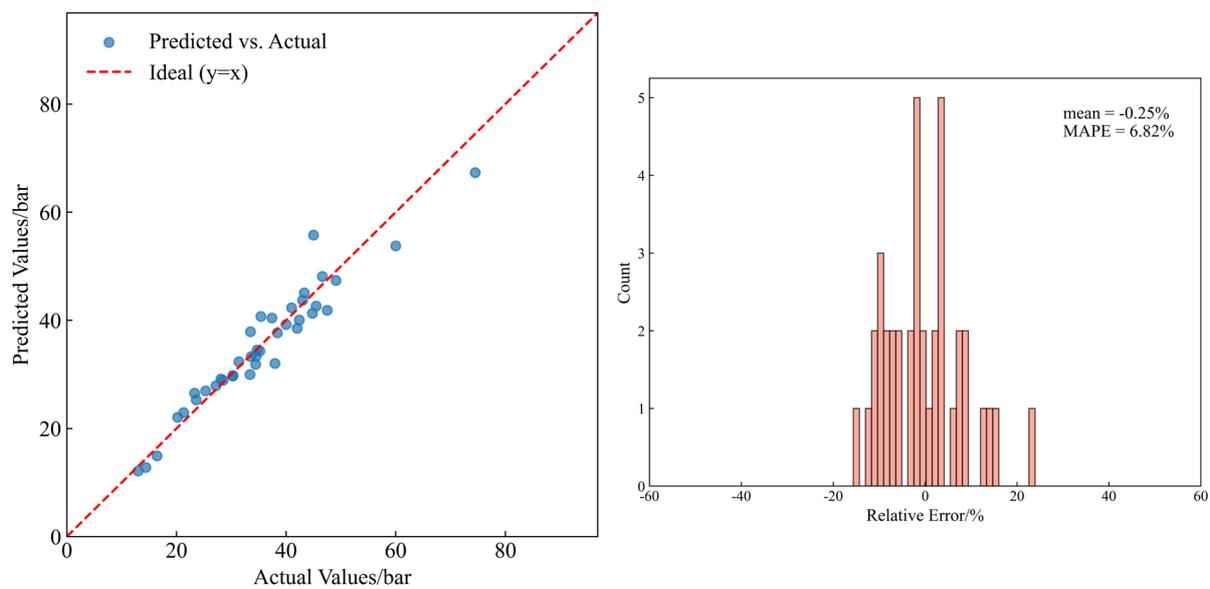

(a) parity plot　　　　　　　(b) relative error distribution

**Fig. S5.** Parity plot and relative error distribution for $P_c$ prediction.

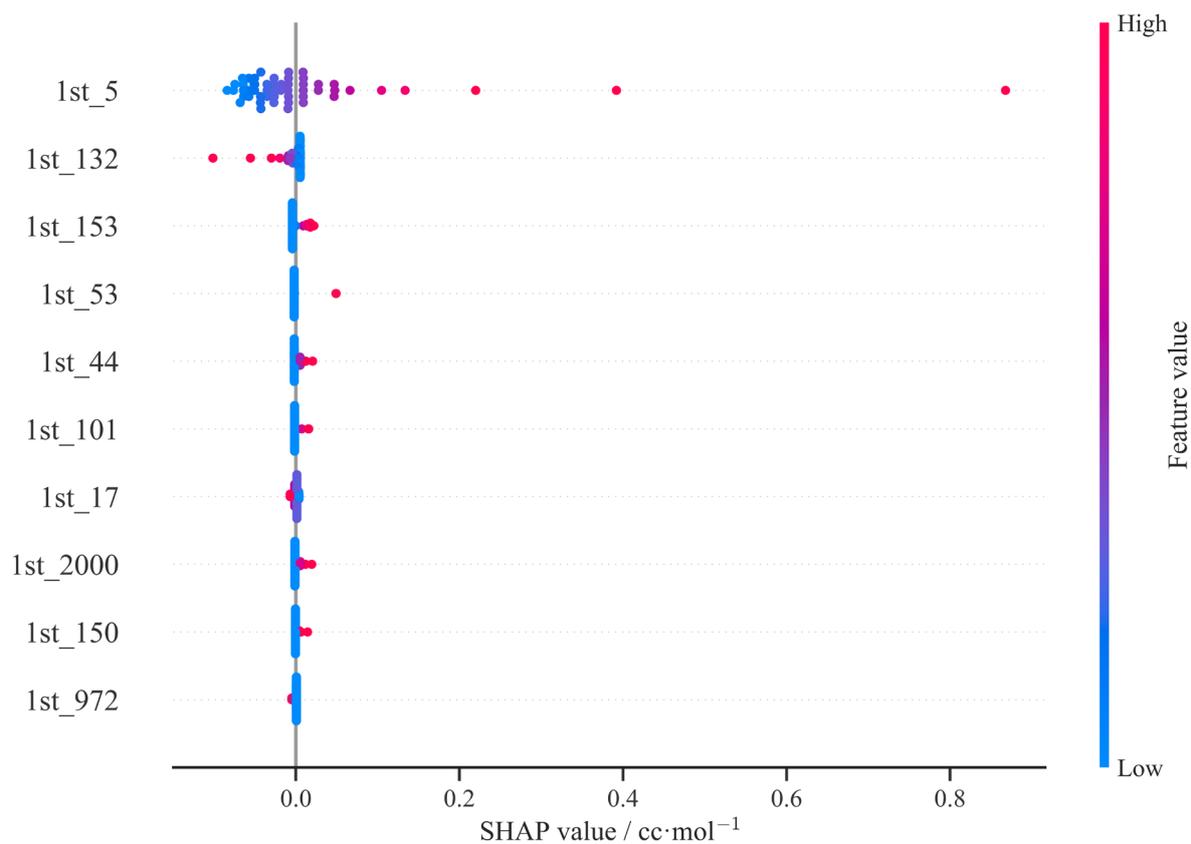

**Fig. S6.** Shapley values of the top 20 features for $L_{mv}$ prediction

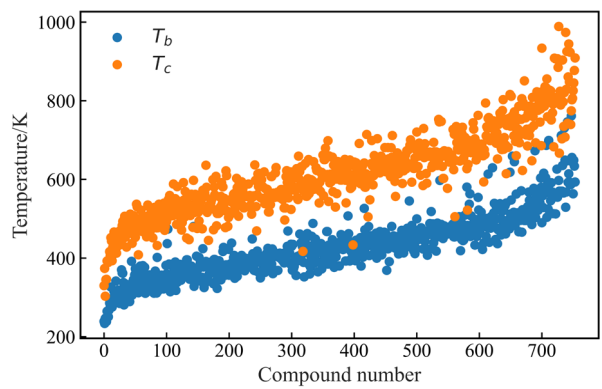

**Fig. S7.** $T_b$ and $T_c$ (y axis) as a function of compound number (x axis)

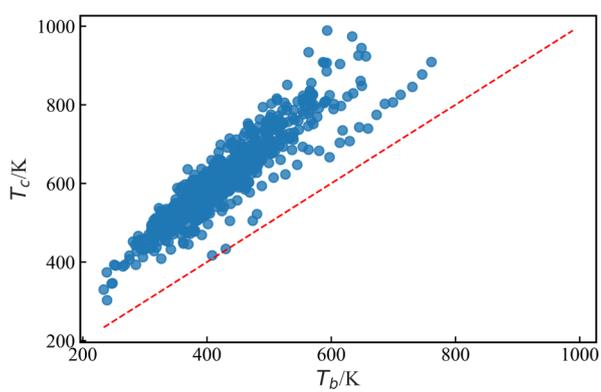

**Fig. S8.** Parity plot of $T_b$ and $T_c$

**Table S1** Six-fold cross-validation results for $P_c$ (feature number =160)

| Fold | RMSE/bar | | | MAPE | | |
|---|---|---|---|---|---|---|
| | Training | Validation | Test | Training | Validation | Test |
| 1 | 3.21 | 4.87 | 4.28 | 0.07 | 0.09 | 0.08 |
| 2 | 2.73 | 5.32 | 3.33 | 0.06 | 0.09 | 0.07 |
| 3 | 3.43 | 3.72 | 3.82 | 0.06 | 0.06 | 0.06 |
| 4 | 3.02 | 5.49 | 4.17 | 0.06 | 0.09 | 0.08 |
| 5 | 3.15 | 4.85 | 3.87 | 0.06 | 0.08 | 0.08 |
| 6 | 3.14 | 4.87 | 3.47 | 0.06 | 0.09 | 0.06 |

**Table S2** Six-fold cross-validation results for $L_{mv}$ (feature number =160)

| Fold | RMSE/cc·mol$^{-1}$ | | | MAPE | | |
|---|---|---|---|---|---|---|
| | Training | Validation | Test | Training | Validation | Test |
| 1 | 0.0084 | 0.0102 | 0.0120 | 0.053 | 0.058 | 0.059 |
| 2 | 0.0084 | 0.0093 | 0.0106 | 0.050 | 0.053 | 0.055 |
| 3 | 0.0080 | 0.0130 | 0.0107 | 0.048 | 0.070 | 0.054 |
| 4 | 0.0087 | 0.0079 | 0.0107 | 0.056 | 0.049 | 0.058 |
| 5 | 0.0079 | 0.0098 | 0.01009 | 0.048 | 0.057 | 0.052 |
| 6 | 0.0084 | 0.0084 | 0.0103 | 0.050 | 0.046 | 0.052 |

**Table S3** Six-fold cross-validation results for $T_c$ (feature number =160)

| Fold | RMSE/K | | | MAPE | | |
|---|---|---|---|---|---|---|
| | Training | Validation | Test | Training | Validation | Test |
| 1 | 38.53 | 56.74 | 40.86 | 0.038 | 0.058 | 0.040 |
| 2 | 33.78 | 40.78 | 42.81 | 0.031 | 0.044 | 0.036 |
| 3 | 38.09 | 37.85 | 42.12 | 0.036 | 0.043 | 0.038 |
| 4 | 37.12 | 46.80 | 40.54 | 0.035 | 0.044 | 0.037 |
| 5 | 41.12 | 43.18 | 45.65 | 0.040 | 0.046 | 0.045 |
| 6 | 30.55 | 53.88 | 41.48 | 0.031 | 0.042 | 0.038 |

**Table S4** 20-fold cross-validation results of GPR-CM for $L_{mv}$ prediction

| Fold | RMSE/ cc·mol$^{-1}$ | | $R^2$ | |
| --- | --- | --- | --- | --- |
| | Training | Test | Training | Test |
| 1 | 0.00064 | 0.0087 | 0.99 | 0.98 |
| 2 | 0.00071 | 0.0095 | 0.99 | 0.98 |
| 3 | 0.00066 | 0.0052 | 0.99 | 0.99 |
| 4 | 0.00064 | 0.0053 | 0.99 | 0.99 |
| 5 | 0.00062 | 0.0088 | 0.99 | 0.94 |
| 6 | 0.00069 | 0.0081 | 0.99 | 0.98 |
| 7 | 0.00065 | 0.0041 | 0.99 | 0.99 |
| 8 | 0.00068 | 0.013 | 0.99 | 0.95 |
| 9 | 0.00068 | 0.0045 | 0.99 | 0.99 |
| **10** | **0.00056** | **0.0031** | **0.99** | **0.99** |
| 11 | 0.00067 | 0.0071 | 0.99 | 0.99 |
| 12 | 0.00066 | 0.0035 | 0.99 | 0.99 |
| 13 | 0.00069 | 0.071 | 0.99 | 0.65 |
| 14 | 0.00067 | 0.0061 | 0.99 | 0.99 |
| 15 | 0.00066 | 0.0073 | 0.99 | 0.98 |
| 16 | 0.00065 | 0.0072 | 0.99 | 0.99 |
| 17 | 0.00060 | 0.0058 | 0.99 | 0.98 |
| 18 | 0.00068 | 0.0064 | 0.99 | 0.98 |
| 19 | 0.00077 | 0.0037 | 0.99 | 0.99 |
| 20 | 0.00076 | 0.013 | 0.99 | 0.97 |

**Table S5** 20-fold cross-validation results of GPR-CM for $P_c$ prediction

| Fold | RMSE/bar | | $R^2$ | |
| --- | --- | --- | --- | --- |
| | Training | Test | Training | Test |
| 1 | 0.56 | 2.92 | 0.99 | 0.90 |
| 2 | 0.36 | 2.12 | 0.99 | 0.95 |
| 3 | 0.56 | 2.52 | 0.99 | 0.94 |
| 4 | 0.60 | 2.30 | 0.99 | 0.94 |
| 5 | 0.61 | 2.36 | 0.99 | 0.95 |
| 6 | 0.45 | 1.95 | 0.99 | 0.94 |
| 7 | 0.58 | 3.01 | 0.99 | 0.95 |
| 8 | 0.54 | 2.98 | 0.99 | 0.94 |
| 9 | 0.52 | 3.57 | 0.99 | 0.85 |
| 10 | 0.55 | 10.43 | 0.99 | 0.52 |
| 11 | 0.48 | 2.72 | 0.99 | 0.95 |
| 12 | 0.43 | 3.52 | 0.99 | 0.94 |
| 13 | 0.59 | 2.69 | 0.99 | 0.91 |
| 14 | 0.55 | 4.46 | 0.99 | 0.81 |
| 15 | 0.53 | 1.55 | 0.99 | 0.97 |
| 16 | 0.55 | 5.88 | 0.99 | 0.76 |
| **17** | **0.54** | **1.06** | **0.99** | **0.99** |
| 18 | 0.39 | 2.52 | 0.99 | 0.93 |
| 19 | 0.46 | 2.56 | 0.99 | 0.93 |
| 20 | 0.50 | 3.27 | 0.99 | 0.83 |

**Table S6** 20-fold cross-validation results of GPR-CM for $T_b$ prediction

| Fold | RMSE/K | | $R^2$ | |
| --- | --- | --- | --- | --- |
| | Training | Test | Training | Test |
| 1 | 7.97 | 25.12 | 0.99 | 0.89 |
| 2 | 7.91 | 26.21 | 0.99 | 0.91 |
| 3 | 7.96 | 27.09 | 0.99 | 0.88 |
| 4 | 7.98 | 25.11 | 0.99 | 0.93 |
| 5 | 7.95 | 36.12 | 0.99 | 0.81 |
| 6 | 2.52 | 55.17 | 0.99 | 0.57 |
| 7 | 7.93 | 25.49 | 0.99 | 0.89 |
| 8 | 7.97 | 31.98 | 0.99 | 0.86 |
| 9 | 7.96 | 32.36 | 0.99 | 0.83 |
| 10 | 7.96 | 29.68 | 0.99 | 0.88 |
| 11 | 7.95 | 29.09 | 0.99 | 0.87 |
| 12 | 7.90 | 23.18 | 0.99 | 0.91 |
| 13 | 2.64 | 67.70 | 0.99 | 0.45 |
| **14** | **7.97** | **20.24** | **0.99** | **0.94** |
| 15 | 7.92 | 27.84 | 0.99 | 0.90 |
| 16 | 7.96 | 21.78 | 0.99 | 0.92 |
| 17 | 7.90 | 119.50 | 0.99 | 0.39 |
| 18 | 7.97 | 22.74 | 0.99 | 0.92 |
| 19 | 7.98 | 27.37 | 0.99 | 0.89 |
| 20 | 7.98 | 35.16 | 0.99 | 0.83 |

**Table S7** 20-fold cross-validation results of GPR-CM for $T_c$ prediction

| Fold | RMSE/K | | $R^2$ | |
| --- | --- | --- | --- | --- |
| | Training | Test | Training | Test |
| 1 | 2.93 | 20.43 | 0.99 | 0.96 |
| 2 | 2.30 | 28.24 | 0.99 | 0.94 |
| 3 | 3.08 | 22.73 | 0.99 | 0.94 |
| 4 | 2.42 | 30.57 | 0.99 | 0.90 |
| 5 | 3.10 | 18.89 | 0.99 | 0.95 |
| 6 | 3.11 | 43.09 | 0.99 | 0.85 |
| 7 | 3.03 | 43.95 | 0.99 | 0.75 |
| 8 | 2.76 | 27.37 | 0.99 | 0.94 |
| 9 | 2.62 | 23.10 | 0.99 | 0.95 |
| 10 | 2.45 | 32.13 | 0.99 | 0.91 |
| 11 | 2.91 | 31.05 | 0.99 | 0.91 |
| 12 | 3.03 | 27.15 | 0.99 | 0.93 |
| 13 | 3.18 | 20.03 | 0.99 | 0.96 |
| 14 | 2.96 | 27.53 | 0.99 | 0.93 |
| **15** | **2.98** | **14.75** | **0.99** | **0.98** |
| 16 | 2.83 | 40.10 | 0.99 | 0.84 |
| 17 | 3.59 | 55.75 | 0.99 | 0.73 |
| 18 | 3.04 | 26.82 | 0.99 | 0.93 |
| 19 | 2.98 | 23.78 | 0.99 | 0.94 |
| 20 | 2.07 | 34.83 | 0.99 | 0.83 |